\documentclass[12pt,preprint]{aastex}

\usepackage{color}

\usepackage[normalem]{ulem}
\usepackage{enumerate}
\usepackage{natbib}
\usepackage{mathtools}
\pdfoutput=1

\newcommand{\pd}{\partial}

\newcommand{\kms}{\,\rm km\,s^{-1} }

\newcommand{\mybf}{}

\usepackage{graphicx}

\begin{document}
\author{A. Dorodnitsyn\altaffilmark{1,2,4}, T. Kallman\altaffilmark{1},  and 
D. Proga\altaffilmark{3}}
\altaffiltext{1}{X-ray Astrophysics Laboratory, Code 662, NASA Goddard Space Flight Center, Greenbelt, 
MD, 20771, USA}
\altaffiltext{2}{Department of Astronomy/CRESST, University of Maryland, College Park, MD 20742, USA}
\altaffiltext{3}{Department of Physics and Astronomy, University of Nevada, Las Vegas, NV 89154, USA}
\altaffiltext{4}{Space Research Institute, Profsoyuznaya st., 84/32, 117997, Moscow, Russia}

\title{
Parsec-scale accretion and winds irradiated by a quasar
}
\begin{abstract}

We present numerical simulations of properties of a parsec-scale torus
exposed to illumination by the central black hole in an active galaxy (AGN). 
{\mybf 
Our physical model allows to investigate the balance between the formation of winds and accretion simultaneously.
Radiation-driven winds are allowed by taking into account radiation pressure due to UV and IR radiation along with X-ray heating and dust sublimation.
Accretion is allowed through angular momentum transport and 
the solution of the equations of radiation hydrodynamics.}
Our methods adopt flux-limited diffusion radiation-hydrodynamics for the dusty, infrared pressure driven part of the flow, along with X-ray heating and cooling. Angular momentum transport in the accreting part of the flow is modeled using effective viscosity.
Our results demonstrate that radiation pressure on dust can play an important role in shaping AGN obscuration. For example,
when the luminosity illuminating the torus exceeds  $L>0.01\,L_{\rm Edd}$,
where $L_{\rm Edd}$ is the Eddington luminosity, we find no episodes of sustained disk accretion because radiation pressure does not allow a disk to form.
Despite the absence of the disk accretion, the flow of gas to smaller radii still proceeds at a rate $10^{-4}-10^{-1}\,M_\odot\,{\rm yr}^{-1}$ through the capturing of the gas from the hot evaporative flow,
thus providing a mechanism to deliver gas from a radiation-pressure dominated 
torus to the inner accretion disk.
As $L/L_{\rm edd}$ increases, larger radiation input leads to larger torus 
aspect ratios and increased obscuration of the central black hole.
We also find the important role of the X-ray heated gas in shaping of the obscuring torus. 
\end{abstract}

\section{Introduction}

{\mybf Active galactic nuclei (AGN) owe their radiative output to accretion of gas which is supplied by the host galaxy.
Many AGNs show evidence for obscuration of the radiation from the central black hole, and many show evidence for
outflows or winds from the central regions.  These phenomena likely arise in the region
$\sim$ 1 pc from the center, where there is a transition between the predominantly cold and dusty
galaxy interstellar medium and the viscous accretion flow onto the black hole.  In this region, the dust from the galaxy
can reprocess the strong radiation field from the center into the infrared (IR), and the relatively
large IR opacity can efficiently couple the gas dynamics to the radiation field.  Numerical models
for dusty accretion flows require treatment of the coupled radiation and gas dynamics in addition to the
effects of radiation on the dust and gas: dust destruction, radiative heating and cooling.
In previous papers we have presented algorithms for treating these processes using simplified assumptions
about the origin of the IR radiation field and also assuming a source of gas on the domain boundary.
In this paper we present models which self-consistently treat the reprocessing of the radiation from the center
into IR, and which have no source of gas outside the computational domain.  These models
demonstrate that  geometrically thick obscuration, outflows, and inflow to the center can all be produced
in a radiation-pressure dominated region near $\sim$1 pc.  



The specific goal of our models include:
\begin{enumerate}
\item 
Our previous work supports the idea that
obscuration in luminous AGN can originate from dusty IR-driven winds.
We  explore this idea further without artificially separating between accretion disk and wind. 
\item
We calculate mass-loss rate and accretion rate. These are time-dependent and
can help to understand the fate of episodic accretion events in luminous AGNs.
\item
Further developing a physical model of the pc-scale AGN yields 
such properties of the wind as  covering factors, column densities, outflow rates,  kinetic luminosities as functions 
of observing angle, and their dependence on accretion rate and luminosity. 
These can provide input for simulations on the larger galactic scale and for models for feedback.
\end{enumerate}

The plan of this paper is as follows:
A discussion of previous work by ourselves and others is presented in Section \ref{Background}.
Also included are simple estimates demonstrating the potential importance of IR radiation radiation pressure;
this section can be skipped by the expert reader.
In Section \ref{Methods} equations of radiation hydrodynamics used to describe the problem are reviewed together with adopted 
approximations, along with the adopted approximations for angular momentum transport and interaction of the radiation with 
the matter. Numerical methods and approximations used to solve the 
system of radiation hydrodynamic equations are summarized in Section \ref{numMethods} and the results are presented in
Section \ref{Results}.

\section{Background}\label{Background}
}

In the AGN region close to the black hole accreting material likely
forms an accretion disk in which angular momentum is transferred via the instability associated with
the distortion of magnetic fields in a conductive shearing flow \citep{BalbusHawley91}.
This occurs within $\sim 10^4 r_g \simeq 10^{16} M_7$\,cm,  where $r_g= 2GM/c^2$, $M$ is the mass of the
black hole, and subscripts indicate scaled quantities, i.e. $M_7=M/10^7M_\odot$.
The inner disk region is responsible for most of the energy generation in AGN but 
in order to reach the inner disk, gas must cross many decades in radius from
regions associated with the host galaxy interstellar medium.  In order to do so, the gas must
lose most of its angular momentum; the details of how this occurs on large scales are likely to be
very different from the behavior in the inner disk itself.   This is due to both the
properties of the cooler gas associated with the
host galaxy, notably the effects of dust, and also to the effect of preheating by the strong radiation from the inner disk.
In this paper, we present models for preheated dusty accretion flows and examine the
behavior of the material as a function of the intensity of the illumination from the inner disk.

Radiative preheating can heat accreting material to temperatures greater than the
local virial temperature.  For gas pressure dominated material, this temperature
is $T_{vir,g} \simeq 10^6 M_7/ r_{\rm pc}$K, where $r_{\rm pc}$ is the distance in parsecs.
If so, the heated material can be levitated into a quasi-static geometrically thick
structure, or can flow outward in a wind.
Such structures have implications for AGN unification, i.e. the hypothesis that the properties of both type I and 
type II AGNs can be explained if the two are intrinsically similar but are viewed from different angles
\citep{Rowan-Robinson77,Antonucci84,AntonucciMiller1985}.
The crucial feature is an optically- and geometrically-thick obscuring structure which prevents
a direct view of the inner disk and black hole from many directions \citep{AntonucciMiller1985}.
Outflows also can couple the properties of the black hole, i.e. its luminosity, with the
properties of the host galaxy, since the outflows can deposit significant energy on galaxy scales.  Also
outflows can limit or regulate the net accretion rate onto the black hole.
The importance of radiation as a regulating agent has been explored in by, e.g. \cite{NovakOstriker2012,Proga_etal08, Proga07}
and its connection to feedback has been demonstrated by \citep[e.g.][]{SilkRees98, King05, Ostriker10,CiottiOstriker07}.

Winds can play a crucial role in the observational appearance of AGNs.
Winds are produced in luminous AGN in the form of UV broad absorption lines (BALs)\citep{Trump06},
UV warm absorbers \citep{Crenshaw97}, and X-ray warm absorbers 
\citep{Reeves03, Blustin03, Gibson09,Gofford11,Page11}.
Evidence for outflows is also seen in the optical spectra from distant
obscured quasars \citep{Greenezakamska12}.   These flows can be 
important to the AGN mass budget, and their rich spectra can provide detailed information about the
outflowing gas:  its ionization, temperature, speed, geometrical distribution, and elemental abundances.
\citep{Murr05}

In dynamical models light from the central engine is obscured by the flows which are either 
inflowing (accretion flows) or outflowing (winds).
Dynamical models for obscuration include: (i) Quasi-stationary models
in which the obscuration is a torus supported by rotation in the radial direction.
and by turbulent motion of molecular clouds in the vertical direction.
Collisions between clouds would likely raise the temperature of the clouds to a 
fraction of the virial temperature $T_{\rm vir,g}\simeq 10^5$K, which is too high to account for observed molecular and 
dust emission.  Magnetic fields have been suggested as a mechanism to provide elasticity to the clouds and avoid
dissipation \citep{KrolikBegelman88,BeckertDuschl04,Nenkova08}. 
(ii) Various mechanisms, including irradiation \citep[i.e.][]{Foulkes10},
can potentially produce global warps in a locally geometrically thin accretion disk
\citep{Phinney89, Sanders89, Pringle92}.  
Such disks can potentially provide obscuration for a significant fraction of lines of sight.
(iii) Geometrically thick flows can be associated with a global magnetic field which allows 
a magnetohydrodynamic (MHD) wind \citep{KoniglKartje94,ElitzurShlosman06} or directly supports a quasi-static 
torus \citep{Lovelace98}, or a combination of radiative and magnetic driving \citep{Emmering92,Everett05, 
  Keating12,KoniglKartje94}.

Dust can play a key role in the properties of accreting or outflowing material near $\sim$1 pc.
Dust survival in the presence of irradiation from the black hole and inner disk depends on the equilibrium
temperature determined by balancing blackbody cooling with the radiative heating by radiation
from the inner disk.  This predicts that dust can survive outside of the sublimation radius 
$R_{\text{sub}} \simeq 0.54 \,T_{1500}^{-2}L_{46}{}^{1/2}\,\text{pc}\mbox{,}$
where the dust sublimation temperature is $T_{1500}=T/1500$K, $L_{46}=L/10^{46}\,{\rm erg~s^{-1}}$,
and $L$ is the radiative luminosity of the inner disk.
Evidence for dust in AGN obscuration comes from 
observations of several nearby AGN where the nucleus is resolved by interferometric techniques.
For example, in NGC 1068 such observations reveal a multi-component,  multi-temperature dusty medium: 
an inner, relatively small ($\simeq 1$ pc) and 
hotter ($\simeq 800$ K) component embedded in a larger ($\simeq 3.5$ pc), cooler ($T\simeq 320$ K) component
\citep{Jaffe2004, Raban09}.  In the Circinus galaxy the 
observed elongated, 0.4 pc diameter component is interpreted as a disk-like structure 
seen almost edge on. The disk-like structure is coincident with that inferred 
from the VLBI maps of $\rm H_{2}O$ maser emission \citep{Greenhill03},
being embedded in a much larger rounded component. This is interpreted as a geometrically thick 
torus with temperature $T\lesssim 300$K \citep{Tristram07}. 
Dust affects the dynamics via its effects on radiation.
Dust opacity, $\kappa_{\rm d}$, in the infrared (IR) is typically  $\simeq 10-30\times \kappa_{e} $  \citep{Semenov03},
where $\kappa_{e}=0.4\,{\rm cm^2\,g^{-1}}$ is the electron scattering opacity.
The critical luminosity at which IR radiation becomes dynamically important is obtained by equating the
dust radiation pressure per H atom with the gravitational force:
$L_{\rm c}=4 \pi c G M_{\rm BH}/\kappa_d \simeq (0.03 - 0.1)\,L_{\rm Edd}$, where
Eddington luminosity, $L_{\rm Edd}$ is the corresponding quantity for
material dominated by the electron scattering opacity:
$L_{\rm Edd}=4 \pi c G M_{\rm BH}/\kappa_e= 1.25 \times 10^{45} M_7\mbox{.}$

The high opacity of dust can effectively trap IR radiation, if it is generated within an optically thick
medium.  In the torus region, the energy density of trapped radiation can exceed the thermal energy density, and we can
define a dusty virial temperature by equating the internal radiation energy density
with the gravitational energy:
$T_{\rm vir, r}\simeq 987\,n_7^{1/4} M_7^{1/4}r_{\rm pc}^{-1/4} \quad{\rm K}\mbox{,}$, where $n$ is the
gas number density.  When the temperature of trapped IR exceeds this value, the material can be levitated, prevented from
accreting, or it can flow out in a wind.
This can  produce a static torus 
\citep{Krolik07,ShiKrolik08} or an outflow if the radiation temperature in the torus exceeds $T_{\rm vir, r}$.

AGN fueling depends on the action of angular momentum 
loss processes, and no single mechanism is likely to 
be dominant over the range of length scales from galactic (kpc) 
down to the black hole event horizon.
On galactic scales, angular momentum can be transferred by
long-range non-axisymmetric gravitational torques acting in the stellar 
and gas system\citep[e.g.][]{LyndenBellKalnajs72}. 
A particular implementation of this effect has been
proposed by \cite{Shlosman89,Shlosman90}, who suggested that
from a typical galactic scale of 10 kpc down to 10 pc, gas accretes via a ''bar 
within bar'' instability: successive bar-type instabilities in a gaseous, self-gravitating disk.
Numerical simulations \citep[e.g.][]{Hopkins12}  give support
for the idea that 
torques generated by non-axisymmetric self-gravitating instability
are indeed responsible for the transport of the angular momentum from 
smaller to larger radii. 

In a self-gravitating environment of the outskirts of an AGN,  a similar effect can operate due to
non-axisymmetric self-gravitating instabilities in a dense gas system. 
If the gas cools faster than the dynamical time-scale $\simeq\Omega^{-1}$, where $\Omega$ is the angular frequency, 
then the disc fragments into individual self-gravitating clumps.
On the other hand, if these two time-scales are comparable,  the disk settles into a gravito-turbulent state \citep[i.e.][]
{Paczynski78,Gammie01}. The non-linear development of self-gravitating instabilities provide dissipation and stresses for the angular momentum 
transport in a gravito-turbulent state . 
At radii $r\gtrsim 1$pc we  would generally expect magnetic field to be less important and self-gravity to be responsible for the angular momentum transport and accretion \citep{Wada12,WadaNorman01}.
At  smaller radii the role of self-gravity diminishes, 
and magnetic stresses become responsible for angular momentum transport and accretion \citep{BKRuz74, 
Narayanetal03, Igumenshchev08}, including the regime where both mechanisms can be 
important \citep{Fromang04a}. The thick torus morphology of \citep{Wada12} was found by \cite{Schartmann2014} to be in a reasonably good agreement of the observations.

The stability of a geometrically, thin self-gravitating disk 
can be described by the Toomre parameter \citep{Toomre64} :	
$Q=\frac{c_s\kappa}{\pi\Sigma G}\,$
where $\Sigma$ is the disk surface mass density, $c_s$ is the sound speed and 
$\kappa=2 \Omega \, r^{-1} \frac{d l}{dr}$ is the epicyclic frequency, and $l$ is the specific angular momentum.
It is usually assumed that a quasi-stationary, self-gravitating accretion disk
hovers on the border of instability maintaining marginal stability and providing $Q\simeq 1$ \citep{Gammie01, 
Goodman03, Rafikov09}. 

The dynamical model of AGN obscuration in which the torus consists of a radiation driven dusty outflow was 
explored in \cite{Dorodnitsyn11a,Dorodnitsyn12a, Dorodnitsyn12b} (hereafter Papers I-III, respectively).
Numerical 2.5D (3D axially symmetric) radiation hydrodynamics  simulations
showed that the reprocessing of the UV and X-ray radiation into IR and the corresponding IR radiation 
pressure on dust grains can drive a strong wind. 
In Papers I and II we calculated the structure of the radiation-driven wind assuming that a certain fraction of the 
incident UV and X-ray radiation are converted into IR.
Thus we did not follow the conversion of X-rays and UV into the IR. Instead, the IR radiation field was assumed 
to enter through the boundary of the computational domain. Flux-limited 
radiation-hydrodynamics was invoked to solve for the radiation-driven wind, originating from an accretion disk. 
The disk at the equator was treated as a boundary condition, 
serving as a source of matter for the wind. 
In Paper III we calculated the conversion of external X-rays into IR, preserving the equatorial symmetry 
and the disk boundary conditions. 
Mass loss rates were found to be in the range $\langle {\dot M} \rangle  \simeq 0.1 - 1.5 M_{\odot}\, {\rm yr}^{-1}$.
This can be compared with the Eddington accretion rate, calculated taking into account electron scattering:
${\dot M}_{\rm E} = 0.02 \,(M_{\rm BH}/10^7 M_{\odot}) M_{\odot}\, {\rm yr}^{-1}$, which suggests that 
it is difficult or impossible to sustain accretion when the gas is affected by such intense radiation pressure.
Accretion with such intense winds should be intermittent with strongly varying accretion rate.


\section{Methods}\label{Methods}

\subsection{Equations of Gas Dynamics and Radiation Transport}

Our physical model describes the time-evolution of a three-dimensional distribution of gas and dust in the 
gravitational field of a supermassive BH, adopting radiation hydrodynamics in 
axial symmetry.   Evolution of dusty gas exposed to IR radiation is described by equations of radiation hydrodynamics. To  first order in $v/c$, these can be written \citep[e.g.][]{MihalasBookRadHydro}:

\begin{eqnarray}
D_{t}\rho &+& \rho\,{\bf \nabla\cdot v}=0\mbox{,}\label{eq11}\\
\rho D_{t}\bf {v} &=& -\nabla p + {\bf g_{\rm rad}} -\nabla \phi  + \nabla \cdot  {\bf t} \mbox{,}\label{eq12}\\
\rho D_{t}\left(\frac{e}{\rho}\right) & = & - p{\bf \nabla\cdot v} - 4\pi\chi_{\rm P}B + c\chi_{\rm E}E
+n^{2} (H_{\rm X}-\Lambda)+\Phi_v
\mbox{,}\label{eq13}\\
\rho D_{t}\left(\frac{E}{\rho}\right) & = & -{\bf \nabla\cdot F} - {\bf \nabla v : P} + 4\pi\chi_{\rm P}B- c\chi_{\rm E}E
\mbox{,}\label{eq14}
\end{eqnarray}
where $D_{t} = \frac{\partial}{\partial t} + {\bf v\cdot \nabla}$ is the convective derivative,
$\rho$ is the mass density and $n$ is the number density of the gas; ${\bf v}$ is the velocity, $p$, and $e$ are gas 
pressure and gas energy density,
$T$ is the gas temperature. We assume a polytropic equation of state: $p=K\rho^\gamma$ for the gas, and $p = (\gamma-1) e$; throughout this work we assume $
\gamma = 5/3$. 

Radiation input from X-ray and UV illumination is taken into account by the $n^2 (H_{\rm X}-\Lambda)$ term in the 
equation
(\ref{eq13}), where $H_{\rm X}$ and $\Lambda$ are respectively heating and cooling functions.
IR~radiation field is described by the radiation energy density, $E$;
radiation flux, ${\bf F}$; and by the radiation pressure tensor, ${\bf P}$; 
opacities: $\chi_{\rm P} (\rm cm^{-1})$, $\chi_{\rm E} (\rm cm^{-1})$ are the Planck
mean and energy mean absorption opacities. Opacities as well as other dependent variables are 
evaluated in the co-moving frame of reference; other radiation-related notation include  the Planck function $B=\sigma 
T^{4}/\pi$, where 
$\sigma = a c/4$ is the Stefan-Boltzmann constant and $c$ is the speed of light; ${\bf \nabla v : P}$ denotes
the contraction $(\partial_{j} v_{i}) P^{ij}$, where $\partial_{j} = \partial/\partial x_j$.
The validity of equations (\ref{eq11})-(\ref{eq14}) in different regimes of the radiation-matter interaction is discussed in 
Paper II. 

The Radiation pressure ${\bf g_{\rm rad} }$ combines pressure of continuum IR radiation on dust, and 
the UV pressure in spectral lines and the radiation pressure due to Thomson scattering on free electrons.
Gravitational forces are assumed to be only due to a supermassive black hole, $\phi = -G M_{\rm BH}/r$. Self-gravity 
of the gas is not taken into account.
Angular momentum transport is allowed by an anomalous viscous stress-tensor, ${\bf t}$, and
viscous dissipation taken into account via a dissipation function $\Phi_v$.

Following previous studies of gaseous, self-gravitating disks \citep{LinPringle87}
we describe the mechanism of angular momentum transport through the introduction of the anomalous viscous stress 
tensor.   The anomalous stress tensor Cartesian coordinates reads

\begin{equation}
t_{ik}=\mu \left(\partial _kv_i+\partial _iv_k-2/3\, \delta _{ik}\, \partial _sv_s\right)
\label{tij1}
\mbox{,}
\end{equation}
where $\mu=\rho\nu _{\rm eff}$, and  $\nu _{\rm eff}\, (\text{cm}^2 \,{\rm s}^{-1})$ is 
the effective kinematic viscosity.

The viscous force is $f_{{\rm vis,i}}=\pd_{k}\,t_{ik}$, where
the summation over a set of similarly indexed terms is implied.
Notice that in case of a geometrically thin equatorial part of the accretion disc, the viscous force simplifies 
to $f_{{\rm vis}, \phi }\simeq  -t_{\text{r$\phi $}}=3/2 \mu  \Omega _K$. 
The viscous heating function, $\Phi_v$ that appears in (\ref{eq13}) reads

\begin{equation}
\Phi_v = \mu \,t_{ik} \pd_k v_i 
\mbox{.}
\label{Phivisc1}
\end{equation}
In our numerical calculations we adopt the 
representation of (\ref{tij1}) and  (\ref{Phivisc1}) in cylindrical coordinates
 \citep[e.g.][]{Tassoul78}.

The effective viscosity,
\begin{equation}
\nu _{\rm eff}=\mathcal{L}_{\rm cr}^2\Omega\mbox{,}
\label{eff_visc}
\end{equation}
depends on the critical spatial scale, $\mathcal{L}_{\rm cr}$ which defines the size of the unstable perturbation.
The latter can be stabilized by the shear, giving $\mathcal{L}_{\rm cr}=G \Sigma/ \Omega^2$, or the perturbation is stabilized by 
the effective pressure, i.e. the velocity dispersion, giving the Jeans scale
$\mathcal{L}_{\rm J}= \sigma^2 / (G \Sigma)$, where $\sigma$ is the local dispersion velocity, \cite{LinPringle87} argue that 
$\mathcal{L}_{\rm J}<\mathcal{L}<\mathcal{L}_{\rm cr}$;
here, for simplicity we assume that $\mathcal{L}\simeq \mathcal{L}_{\rm J}$. When performing numerical simulations (see further in the 
text) we experimented with $\mathcal{L}\simeq \mathcal{L}_{\rm cr}$ and did not find any noticeable difference in the results.

\subsection{Photoionization equilibrium and X-ray heating and cooling}

The radiation field in our calculation is divided into the incident continuum
from the inner disk and black hole, and the diffuse IR produced by thermal
emission within the accretion flow.
The incident continuum is treated using a single stream
 approximation. We trace X-rays from a central point source and  
assume attenuation according to $F_X \simeq \exp(-\tau_X)$, where the optical depth with respect to X-ray absorption 
is
${\tau_{\rm x} = \int \, \kappa_{\rm x}\rho \,dr }$, and  $\kappa_{\rm x}$ is the X-ray opacity.

We assume that the ionization state of the gas can be parametrized by ionization parameter,
$\xi$:

\begin{equation}\label{smallxi}
\xi=4\,\pi\,F_{\rm x}/n \simeq 4\cdot 10^3 \cdot f_{\rm x}\,\Gamma\, M_7/ (N_{23}\,r_{\rm pc})
\mbox{,}
\end{equation}
where $N_{23}$ is the column density in $10^{23}$ ${\rm cm}^{-2}$; 
$F_{\rm x} = L_{\rm X}/4\pi r^2$ is the local X-ray  flux; X-ray luminosity of the nucleus:
$L_{\rm X} = f_{\rm x} L_{\rm accr}$, where
$L_{\rm accr}=\Gamma L_{\rm edd}$ is the total accretion luminosity. This relation also serves as a definition of the model parameter $\Gamma = L_{\rm accr}/L_{\rm edd}$
where $L_{\rm edd}=1.25\cdot 10^{45}\,M_7$ is the Eddington luminosity, and  $M_{7}$ is the mass of a BH in $10^{7} M_{\odot}$.

Energy deposition by external X-rays is calculated making use of heating and cooling functions. These include 
Compton heating and cooling:

{\mybf
\begin{equation}\label{heating1}
H_{\rm IC}({\rm erg\,cm^{3}\, s^{-1}})= 8.9\cdot 10^{-36}\,\xi\,(T_{\rm x}-4T)\mbox{,}
\end{equation}

photo-ionization heating-recombination cooling:

\begin{equation}\label{heating2}
H_{\rm PI}({\rm erg\,cm^{3}\, s^{-1}})= 1.5\cdot 10^{-21}\,\xi^{1/4}\,T^{-1/2}(T_{\rm x}-T)T_{\rm x}^{-1}\mbox{,}
\end{equation} 

\noindent {\mybf (and $H_{\rm X}=H_{\rm PI}+H_{\rm IC}$),} bremsstrahlung and line cooling:

\begin{eqnarray}\label{heating3}
\Lambda({\rm erg\,cm^{3}\, s^{-1}})&=&3.3\cdot 10^{-27} T^{1/2}\nonumber\\
&+&(4.6\cdot 10^{-17} \exp(-1.3\cdot 10^5/T)\xi^{(-0.8-0.98\alpha)} T^{-1/2}+ 10^{-24})\, 
\delta\mbox{.}
\end{eqnarray}
We take the parameter $\delta\simeq 1$ representing optically thin line cooling.

Heating and cooling rates 
$H_{\rm IC}$, $H_{\rm PI}$ and $\Lambda$ are calculated making use of the XSTAR photo-ionization code \citep{KallmanBautista01}
assuming ionizing continuum with (energy) power law index of $\alpha$=1.}
Results were approximated to a reasonable ($\simeq 25\%$) accuracy
by the analytical formulae, originally derived by \cite{Blondin94} for a 10 keV bremsstrahlung spectrum with $T_{\rm 
X}=2.6\times 10^7$K.
Formulae (\ref{heating1})-(\ref{heating3}) are the slightly improved version by \citet{Dorodnitsyn08b} which 
incorporate newer atomic data.

\subsection{Cold gas and dust}\label{gasanddust}
Interaction of X-ray radiation with the cold molecular and dusty gas depends on various physical and chemical 
processes \citep{Maloney96,KrolikLepp89}.
In general, the proper incorporation of this material calls for consideration of all the radiative, chemical and dust 
network of processes. 
To couple this to the radiation hydrodynamics we adopt a simplified treatment which 
 nevertheless preserves the qualitative behavior of gas and dust in the presence of an intense 
radiation field. 
We separate the interaction into different regimes depending
on whether the gas is highly ionized, primarily atomic, or largely molecular. To decide between different regimes we 
check if an effective ionization parameter meets certain threshold requirements. Thus,
following \citep{Maloney96}, cooling of the cold gas depends on the value of the effective ionization parameter, 
$\xi_{\rm eff} =\xi/N_{22}^{0.9} = 1.26\times 10^{-1} F_{\rm x}/ (n_{8} N_{22}^{0.9})$
which is a switch that determines the cooling channel and whether the gas is assumed to be primarily in atomic or 
molecular state. For details of how we treat this problem, refer to Paper III.  

\subsubsection{Dust and gas coupling}

Throughout this work we treat dust and gas as one fluid.  This is a simplification that implies both momentum and positional coupling.  
Momentum coupling means that the momentum of the accelerated dust grain is effectively shared with the gas molecules and ions; the positional coupling means that the grain doesn't travel far alone with respect to the gas. 
On the other hand, momentum coupling and no positioning coupling are the usual assumptions in the calculations of dusty radiation-driven winds, such as winds from AGB stars
\citep{Gilman72, LamersCassinelli99}.
The conditions, most notably, the characteristic length scales, in our wind are different from those of a wind from a star.
In the following and throughout our calculations, we adopt the following fiducial values for the parameters of  the dust: the grain radius, $r_{\rm d}=1\times 10^{-4}\,{\rm cm} = 1 \mu$, the density of the dust grain $\rho_{\rm d}=\rho_{\rm d,1} =1\, {\rm g\, cm^{-3}}$,  and the dust-to-gas mass ratio $f_{\rm d}=1\times 10^{-2}$. Above the sublimation temperature, $T_{\rm sub}=1500\,{\rm K}$ the dust is assumed to be destroyed.  
The assumption of dust and gas coupling is justified for the conditions relevant to our simulations, as can be shown via simple estimates.

The analysis, given by \cite{Krumholz13}  for the dust grain collisional stopping length, gives
$\l_\text{gr}\simeq 7.25\times 10^{-7}\, \frac{n}{10^8}$pc \citep[see also ][]{Murr05}. This length scale is a factor $10^{-4}$ smaller than the smallest length scale in our simulations.
Sticking of  impinging electrons and ions and interaction with intense X-ray and UV radiation makes the grains to possess a charge which further couples them to gas \citep{DraineSalpeter79}.
If there is no significant magnetic field, then the interaction between grains and ions is governed by the grain stopping time which  can be estimated taking into account the fact that dust grains are much heavier than ions. If so \cite{Spitzer1978Book} gives $t_{\rm s} \propto {r_{\rm d} \rho_{\rm d}}/{\rho v_{\rm th}} \simeq 3.6\times 10^{-2} (\frac{n}{10^8})^{-1}(\frac{T}{500})^{-1/2}$yrs where $v_{\rm th}$ is the thermal velocity of gas. Comparing it with the sound crossing time we have: $t_{\rm s} / t_{ \rm cross} \simeq 
7.45\times 10^{-8}(\frac{n}{10^8} \frac{r}{1pc} )^{-1}$.

A different estimate comes from that AGN torus region should have small scale magnetic fields (i.e. see the discussion in \citet{KrolikBegelman88}). 
On the microphysical level, the gas at the torus region should be turbulent, with small-scale magnetic field of the order of the equipartition level. 
In this case the grain stopping time is defined by 
$\min(l_{\rm gr}, r_B)$, where $r_B$ is the gyroradius. 
The dust grain that possesses a charge, $Z$ will gyrate in the magnetic field ${\bf B}$ with the gyration radius: $r_B \sim v_d/(Z e B)$.
The small-scale equipartition magnetic field 
$B_{\rm eq} \simeq 4 \, (\frac{n}{10^8} \frac{T}{500} \, {\rm K} )^{1/2} {\rm \mu G} $ thus gives $r_B \simeq 2 \times 10^{-4} \frac{v}{10} \left(\frac{n}{10^8}\frac{T}{500}\right)^{-1/2}$pc for $Z\simeq 20$. 
Thus the gyroradius is small compared with the size of the region, and compared with our numerical resolution.  Dust particles must move with the small scale field, and hence with the entire plasma, on length scales $\geq r_B$.

These estimates show that $l_{\rm gr}/r \ll 1$ and the dust-grain mixture can be treated as a single fluid. The assumption of the positional coupling can break at high altitude, i.e. close to the torus funnel where the density is very small. However, the intense X-ray radiation of AGN makes an important difference between the torus winds and a dusty stellar wind. Our numerical results show, a posteriori, that no dynamically significant amounts of dust survive 
in the region of low density because of the strong X-ray radiation there.
Though it would be interesting to consider the effects of dust separation at high altitudes  this is beyond the scope of the current work.

\subsubsection{Dust and molecular phase}
If the ionization parameter is  small
$(\xi_{\rm eff} \ll  \xi_{\rm m} =10^{-3})$ the gas is primarily molecular. 
When $\xi_{\rm eff} <  \xi_{\rm m}$, 
we assume radiative equilibrium between molecular gas and dust, i.e. $T_{\rm g} = T_{\rm d}$. The equilibrium dust 
temperature is found 
assuming that dust can directly reprocess X-rays to IR (see Paper III).
If $\xi_{\rm eff} \geqslant \xi_{\rm m}$ and $T<T_{\rm sub}$ we assume that the exchange between the radiation  temperature, $T_{\rm r}(E)$ and $T(e)$ is determined from 
equations (\ref{eq13})-(\ref{eq14}). The gas temperature $T$ can contribute to IR and
vice versa.  In equation (\ref{eq14}) we assume that ${\bf F} = {\bf F_{\rm IR}}$, where  ${\bf F_{\rm IR}}$ is the  infrared flux. The details of how there radiation pressure is calculated are given in Section \ref{SectRadPressure1}.
For $T>T_{\rm sub}$ we assume that the dust is destroyed. The opacity of the gas, in this case, is taken to be that of Thomson.

\subsection{IR radiation pressure}\label{SectRadPressure1}

Radiation pressure consists of two components: i) pressure of the continuum IR radiation on dust, ${\bf g_{\rm IR}}$ 
ii) Another component of the radiation pressure - the radiation 
pressure in lines, ${\bf g_{\rm l}}$ is due to interaction of the UV radiation with spectral lines (Section 
\ref{linepressure}). The total radiation pressure: 

\begin{equation}\label{radforcetot}
{\bf g_{\rm rad}}	= {\bf g_{\rm c}} + {\bf g_{\rm l}}\mbox{.}
\end{equation}
The continuum radiation pressure: ${\bf g_{\rm c}}={\bf g}_{\rm T}+{\bf g_{\rm IR}}$ where
${\bf g}_{\rm T}= {\bf F} \kappa_{e}/c$ is the radiation pressure due to Thomson scattering
and ${\bf F}$ is the local  radiation flux.  

In order to make the problem tractable in the framework of radiation hydrodynamics we make several simplifications:
We assume that the UV flux $f_{\rm uv}F$ contributes only to the  radiation pressure in spectral lines 
(see next paragraph). Similarly,  X-ray flux $f_{\rm X} F$ contributes to X-ray heating;  Both fluxes are attenuated through a simple, $\propto e^{-\tau}$ prescription. In all our models we assume $f=f_{\rm UV}=f_{\rm X}=0.5$.

\noindent The IR radiation pressure is given by:

\begin{equation}\label{radforce}
{\bf g_{\rm IR}}	= \frac{1}{c}\frac{\bf \chi_{\rm F} \,{\bf F_{\rm IR}} }{\rho}\mbox{,}
\end{equation}
where the total flux mean opacity, $\chi_{\rm F}$ consists of absorption opacity and the Thomson scattering opacity.  
For simplicity, we do not differentiate between $\chi_{\rm F}$, $\chi_{\rm P}$ and $\chi_{\rm E}$.

\noindent The IR radiation flux ${\bf F}_\text{IR}$ that is needed in equations, (\ref{eq14}) and (\ref{radforce}) is calculated 
adopting a Flux Limited Diffusion (FLD) approximation.  The ${\bf F}_\text{IR}$ is calculated from the diffusion law:

\begin{equation}\label{FickLaw}
{\bf F}_\text{IR}= - D \, \nabla E \mbox{.}
\end{equation}
If optical depth is large $\tau\gg 1$, the diffusion coefficient is:
$D= c\, \lambda$.
where $\lambda=1/(\kappa_{\rm d} \rho)$ is the photon mean free path, and $\kappa_{\rm d} = \chi/\rho$ is the dust 
opacity per unit mass.
Notice, that in the diffusion regime $\kappa$ cancels out from the formula for ${\bf g}_{\rm rad}$. 

That is 
if the optical depth of the gas-dust mixture in the IR
$\tau_{\rm d}\gg 1$, radiation pressure ${\bf g}_{\rm rad} \propto  \nabla \,E \propto \nabla \,T_r^4$ where 
$T_r$ is the the radiation temperature. When $\tau_{\rm d}\ll 1$ radiation pressure scales as ${\bf g}_{\rm rad} \propto E$.

The diffusion approximation tacitly assumes that optical depth $\tau \gg 1$.
Most of the torus where IR pressure is important indeed has $\tau_{\rm d}>1$.
When  $\tau \ll 1$ the mean free path, $\lambda \to \infty$,  and
$D\to \infty$, and $|{\bf F}|\to \infty$. 
In a free-streaming limit however, there should be $|{\bf F}|\to c E$.
That is, when optical depth becomes small, or when $\rho\to 0$, the diffusion approximation is no longer applicable.
Thus, at $\tau<1$ it must be modified to obtain the correct limiting behavior
in the radiation free-streaming limit. 

Thus, in order to describe correctly regions of small $\tau$, 
we adopt
the flux-limited diffusion approximation \citep{AlmeWilson74,Minerbo78,LevermorePomraning81}.
In this approximation $\lambda$ is replaced by $\lambda^{*} = \lambda\,\Lambda_{\rm F}$,
where $\Lambda_{\rm F}$ is the flux limiter. The flux limiter we adopted in Paper I,II and in the current work is that of 
\cite{LevermorePomraning81}:

\begin{equation}\label{LP_FluxLim}
\Lambda_{\rm F} = \frac{2+R_{\rm LP}}{6+3R_{\rm LP}+R_{\rm LP}^{2}}\mbox{,}
\end{equation}
where $R_{\rm LP}=\lambda\,|\nabla E |/E$. One can see, that if $\tau\to 0$, then $R_{\rm LP}\to \infty$, and 
$|F| \simeq c\,E$; and if $\tau\gg1$ $R_{\rm LP}\to 0$ and $\Lambda_{\rm F}\to 1/3$.

\subsection{Radiation pressure in spectral lines}\label{linepressure}

Where the gas is hot and partially ionized, we take into account the pressure of UV radiation in spectral lines:  

\begin{equation}\label{radforcelines}
{\bf g}_{\rm l}=( f{\bf F}\kappa/c)\,M(t)\mbox{,}
\end{equation}

where $f F$ (see Section \ref{SectRadPressure1})  is the local UV flux, and 
$M(t)$ is the force multiplier \citep{CAK75,OwoCastRib}:

\begin{equation}
M(t)=k\,t^{-\alpha} 
\left( (1+\tau_{\rm max} )^{(1-\alpha)}
-1 \right)/\tau_{\rm max}^{1-\alpha}\mbox{,} \label{Mt1}
\end{equation}

\noindent where $t=\tau/\eta_l$ is the optical depth parameter,  $\eta_l=\kappa_l/\sigma_e$ is the line strength parameter
(different from the accretion efficiency used earlier), 
$\sigma_e=\kappa_e/m_p$ is the Thomson cross-section, $m_p$ is a proton's mass,
and $\tau_{\rm max}=t\,\eta_{l,\rm max}$;
a parameter $\eta_{l,\rm max}$ was introduced to limit the effect of very 
strong lines \citep{OwoCastRib,StevensKallman}.  
Local UV flux is calculated from the total bolometric luminosity $L$ adopting a simple attenuation model and 
assuming 
the nucleus radiates $f_{\rm UV}L$ in UV. The parameter $k(\xi)$ in (\ref{Mt1}) reads \citep{StevensKallman}:

\begin{equation}
k=0.03+0.385\exp(-1.4\,\xi^{0.6})\mbox{,}
\label{k}
\end{equation}
where $\eta_{\,l,\rm max }(\xi)$ is found from
$$
\log_{10}\eta_{\,l,\max}=\left\{
\begin{array}{ll}
6.9 \exp( 0.16\, \xi^{0.4} )\mbox{,} & \log_{10}  \xi \le 0.5\mbox{,}\\
9.1 \exp(-7.96\cdot 10^{-3} \, \xi) \mbox{,} &\log_{10} \xi > 0.5\mbox{.} \label{eta_xi}
\end{array}
 \right.
$$
As it follows from the results of our numerical simulations,
the large opacity of dusty gas with respect to UV makes ${\bf g_{\rm l}}$ component of 
the acceleration negligible in the dusty component of the flow.

\section{Numerical Methods}\label{numMethods}

The system of radiation-hydrodynamics equations (\ref{eq11})-(\ref{eq14}) is solved with our 
radiation-hydrodynamics code. The hydrodynamic part of this system is solved with the 
methods implemented in the ZEUS-MP \citep{Hayes06} version of the original ZEUS code \citep{Stone92a}.
The radiation-hydrodynamic part is designed and tested in Papers I-III
and conforms with methods and structure of the original ZEUS code \citep{Stone92a}.

We calculate the dynamics throughout the entire volume occupied by the  accreting and outflowing gas.
In this way, we address a variety of possible ways in which an accretion inflow and the wind can co-exist.
The addition of this new capability to treat accretion requires to compute  the $\nabla \cdot {\bf t}$ 
term in the momentum equation, (\ref{eq12}), and the viscous heating term in the 
energy equation,  (\ref{eq13}). During the time-step update this is done adopting an operator-splitting technique.

We solve the time-dependent system of equations (\ref{eq11})-(\ref{eq14}) in cylindrical 
coordinates $\{R, z\}$, adopting a uniform cylindrical grid, $\{R, z \}$ with the $N_z = N_R =300$.  The grid extends from approximately 
the dust sublimation radius,
$R_{\rm in}=0.5\,$pc to $R_{\rm out}=2.5\,$pc in the radial,  and from
$z_{\rm low}= -1\,$pc to $z_{\rm up}=1\,$pc in the vertical direction.
Three components of the velocity, $v_r$, $v_z$, $v_\phi$ are taken into account, 
assuming azimuthal symmetry, $\partial _\phi = 0$, i.e. adopting the $2.5D$ approximation. 
 
Boundary conditions (BC) for ${\bf v}$, $\rho$, $e$ 
at the boundaries of the computational domain are set to be outflowing. The radiative BC are free-streaming $F\simeq 
cE$ at all boundaries. Radiation BC are implemented in a way that preserves second order accuracy, 
required for proper interoperability with other numerical methods in our original  diffusion solver.

Radiation hydrodynamics is a highly non-linear problem, and 
obtaining the numerical solution of the system (\ref{eq11})-(\ref{eq14}) is challenging.
For example, gas temperature depends on X-ray heating which depends on density.
Temperature gradients create strong local radiation pressure, 
providing a feedback loop to dynamics (see Paper II for the description of the radiation time-scales).

The solution of coupled equations, (\ref{eq13})-(\ref{eq14}) for $E$ and $e$ requires the solution of a general fourth order algebraic equation (Paper II). This is very sensitive to the range of parameters of this equation such as time  step, $dt$ etc. 
To make things worse, a strong non-linearity affects many parts of our system: X-ray heating-matter interaction; coupling between  thus obtained gas temperature and radiation energy equation; finally, in the momentum equation there is a strong dependence on the gradients of the radiation energy density, and again, implicitly on the rate of energy deposition from X-ray heating. 
To tackle this problem(s) we often have to split the time-step and then separately
sub-cycle the radiation hydrodynamics part of equations in time until  the desired time-step is achieved. 
A similar line of arguments but perhaps less restrictive applies to coupling between the dynamics and the effects of viscosity.

\subsection{Initial conditions}

Simulations start from a rotating, polytropic, i.e. $p\simeq \rho^{\gamma}$, constant angular momentum torus that is 
imbedded in low-density gas. 
An analytical solution that describes such a torus does exist, \citep{PapaloizouPringle84} but does not take into 
account radiation pressure. 
We calculate the initial distribution of the density from an approximate formula 
which in the non-radiative case reads:

\begin{equation}
\frac{p}{\rho}=\frac{G M (\gamma-1)}{\gamma R_0} \left( 
\left(\frac{G M  (1- \Gamma e^{-\tau_{\rm X}}) }{R^2+z^2}\right)^{1/2} - 
\frac{1}{2} \left( \frac{R_0}{R} \right)^2  - \frac{1}{2d} \right)
\mbox{,}\label{PaPTorus}
\end{equation}
where $R_0$ is the distance to the locus of the density maximum, $\rho_0=m_p n_0$; the parameter $d>0$ 
measures the distortion of the torus's polar cross-section.  
Notice  that in the radiation-dominated case $\gamma=\gamma_{\rm r}=4/3$. 
To calculate this initial distribution we do the following: we calculate $\rho$ from (\ref{PaPTorus}) assuming a polytropic torus with $\gamma=\gamma_{\rm r}=4/3$; substitute in (\ref{PaPTorus}) total pressure $a T^4/3 + \rho {\cal R} T\to p $, where
${\cal R} \simeq 8.31\times 10^7 {\rm erg\, K^{-1}\,g^{-1}}$ is the universal gas constant.
Thus, we assume that the gas and radiation are initially in equilibrium and calculating the distribution of $T$; in the rest of the simulations we take $\gamma=5/3$.

\subsection{Input Parameters}\label{Input}

Radiation energy input into the flow is controlled by the major input parameter $\Gamma=L/L_{\rm Edd}$. We present several models for $\Gamma =\{0.01, 0.05, 0.1, 0.3\}$ which illustrate the range of behaviors spanned by interesting values of $\Gamma$.
Local radiation flux depends on the attenuation and thus on the density scale, $n_0$ of the initial density distribution, 
$(\ref{PaPTorus})$. 
All models have $ n_{0}=1 \times 10^{8} \, {\rm cm^{-3} }$ which corresponds to the initial torus mass, $M_{\rm tor}
=5\times 10^{5}M_{\odot}$.
The boundaries of the computational domain are located at $R_{\rm in}=0.5\,\text{pc}$ and  $R_{\rm out}=2.5\,
\text{pc}$ and remain fixed for all models.
Other parameters are $M_{\rm BH}=1\times 10^{7} M_{\odot}$, $f_{\rm X}=0.5$, and 
$f_{\rm UV}= 0.5$.
The evolution of the radiative flow was followed for $10$ dynamical times, $t_0\simeq  4.7\times 10^{3}\, r_{\rm pc}^{3/2} M_{7}^{-1/2}{\rm yr}$. 



\section{Results}\label{Results}
\subsection{General effect of radiation on a torus  }

Radiation influences gas dynamics in our models in several ways: 
\begin{enumerate}
\item[i)]
X-ray irradiation of the torus creates an overheated thin layer - a 'skin' where the ionization parameter $\xi \geq 1$ and X-rays heat the gas to
a temperature comparable to the local {\it gas} virial temperature, so that it expands into the inner cone-shaped throat or funnel at small radius. 
Eventually, this hot gas completely fills the throat of the torus. Some of this gas is accreted i.e. crosses the left boundary of the computational domain,
and  some flows to larger radii as a thermal wind.  
Confinement by the cold and dense torus creates a funnel flow - a fast outflow of a 
hot and diluted gas within an approximately conical region surrounding the disk axis.
\item[ii)]
UV and X-rays are absorbed in the interior of the torus, and reradiated as thermal IR at the local equilibrium temperature. 
This may increase the internal torus radiation temperature to or above the local {\it radiation} virial temperature. Then the torus expands and the outflow commences.
\item[iii)]
Radiation deposited in the inner disk also exerts a net bulk radiation
pressure on the torus, pushing it outward.  
\item[iv)]
Viscous torques redistribute angular momentum within the torus allowing the gas 
to move inward, i.e. to accrete.  This process is enhanced by radiative heating
of the disk interior.
\item[v)]
Back-reaction from the hot gas on the torus. Hot gas created in step (i) pushes the torus further outward ablating and compressing it in the vertical direction.

\end{enumerate}
These processes  are illustrated in the following results.  We present four models, which differ by their
values of the Eddington ratio $\Gamma$.
Dynamical and obscuring properties of this models are described in 
Sections \ref{AccretionRate}, \ref{MassLossRateAndWindEnergy}, and \ref{ObscuringProperties}.


\subsection{Model with $L = 0.01\, L_{\rm edd}$ }

\begin{figure}[htp]
\includegraphics[width= 450pt]{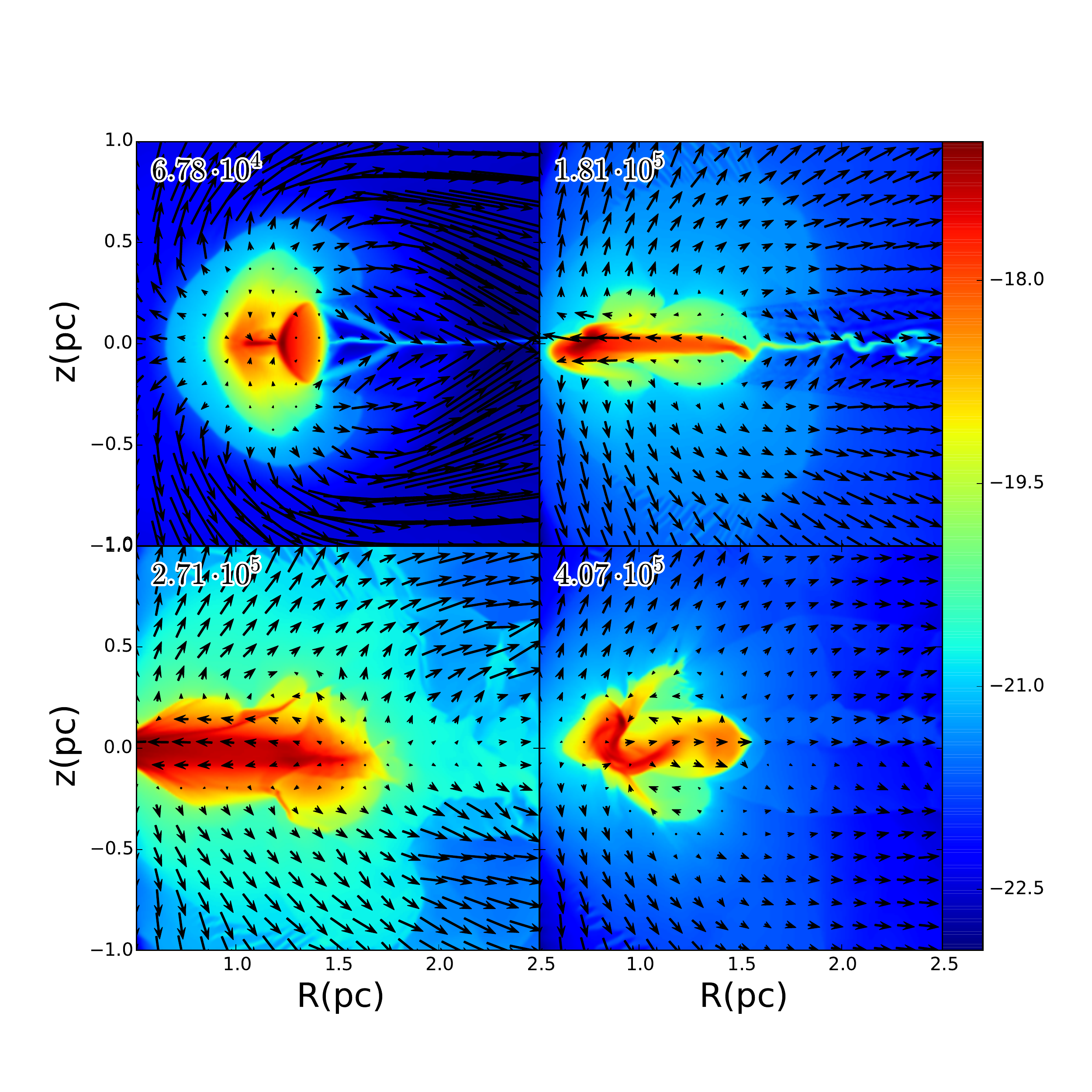}
\caption{
Color plot of the density, $\log\rho$ in ($g\,{\rm cm^{-3}}$) for the model with $L= 0.01 L_{\rm edd}$,
shown at different times given in years.
Axes: $z$: distance from the equatorial plane in parsecs; $R$: distance from the BH in parsecs.
}\label{figGam001}
\end{figure}

At low Eddington ratio the effects of internal radiation pressure are small compared with the
other radiation influences described above.
Figure \ref{figGam001} shows the density map for the model with $L = 0.01\,L_{\rm edd}$.
This model is characterized by two distinct components: a hot evaporative outflow, and a colder thin accretion disk.
Radiation heats the gas in the inner torus ''funnel" through Compton and photo-ionization heating,  raising the  temperature to $10^5-10^6$K.
This occurs within a Compton heating time
$t_{\rm heat}=kT r^2/(\Gamma GMm_Hc \Delta \varepsilon_{\rm C}) \simeq 3.2 \, T_4 r_{\rm pc}^2 \Gamma^{-1}  M_7^{-1}{\rm yr} $,
where $\Delta \varepsilon_{\rm C}$ is the Compton energy transfer per scattering.
This is fast compared with the dynamical time associated with gravitational motion.
Thermal expansion times are also rapid compared with the dynamical time
so that
after $t \simeq 10^3-10^5$ yrs, hot gas evaporated from the torus fills approximately half the domain,
This can be seen at the earliest time-step shown in Figure \ref{figGam001}
and corresponds to step (i) above.  Temperatures within this flow range from 
$10^4$ to a fraction of the Compton temperature, $T_{\rm X}=2.6\times 10^7$K.
An outflow occurs mostly in a form of evaporative flow with $T\simeq 10^5-10^6$K.
This gas leaves the computational domain as a wind. For example, the maximum velocities at the $z-$ and $r-$ directions:
$v_{z,\rm max}=240\,{\rm km\,s^{-1}}$, at $R\simeq 0.67$ pc and $v_{r,\rm max}=220\,{\rm km\,s^{-1}}$ at $R\simeq 2.2$ pc.

The influence of infrared radiation pressure on the flow depends on the value of $\Gamma$ via the dust
effective temperature.  At low $\Gamma$, the infrared radiation input into the interior of the disk is
not strong enough to generate any significant radiation pressure on dust.
This can be seen from Figure \ref{figGam001}
at times greater than $\sim 10^5$ yrs:  the cold core of the thin disk is not disrupted rapidly.
Consequently, there is no infrared-driven wind formed.
Inside the torus, viscous torques redistribute angular momentum of the gas allowing the gas 
to move inward, i.e. to accrete.   
The important characteristic of this model is that it shows a significant period of disk accretion.
This is apparent at time $\sim 10^5$ yrs; a tail of higher angular momentum material moving out at
large radii is a signature of accretion.  This structure is also compressed by wind material.

At later times, $t \simeq 3\times 10^5$yrs
evaporation from the disk significantly depletes the disk mass.
This reduces the net inflow, so that evaporation from the inner disk edge ultimately stops the accretion
of disk material. 
The disk is truncated and hot, much more compact but still geometrically thick torus which
appears roughly at the same place where the initial torus was located. The mass of this torus is much less than
the initial mass $M_{\rm tor}(t=4.07 \times 10^5 \,{\rm yrs})\simeq 1.4\times 10^4\,M_\odot$.

\begin{figure}[htp]
\includegraphics[width= 450pt]{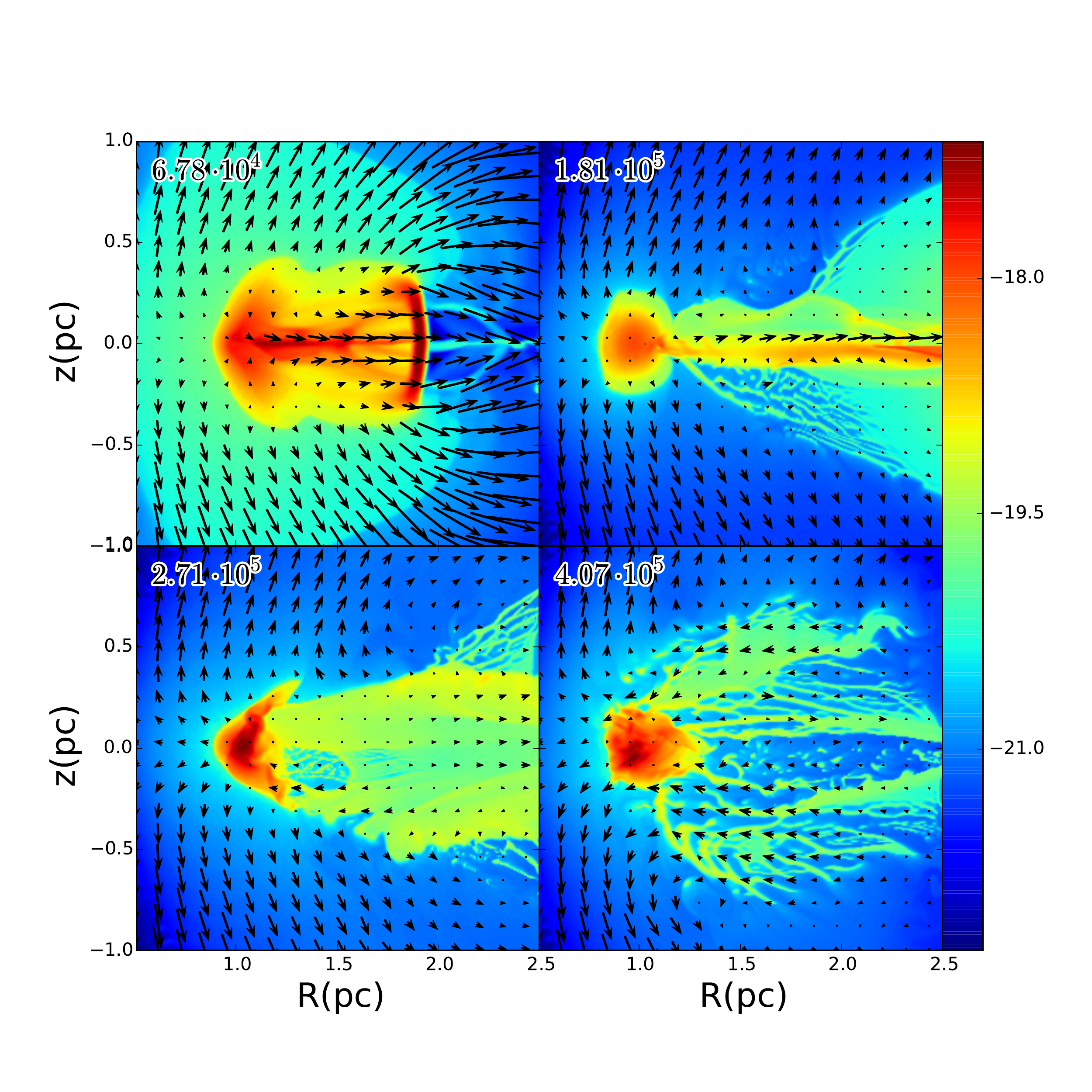}
\caption{
As in Figure 1 but for
$L= 0.05 L_{\rm edd}$.
}\label{figGam005}
\end{figure}
\begin{figure}[htp]
\includegraphics[width=450pt]{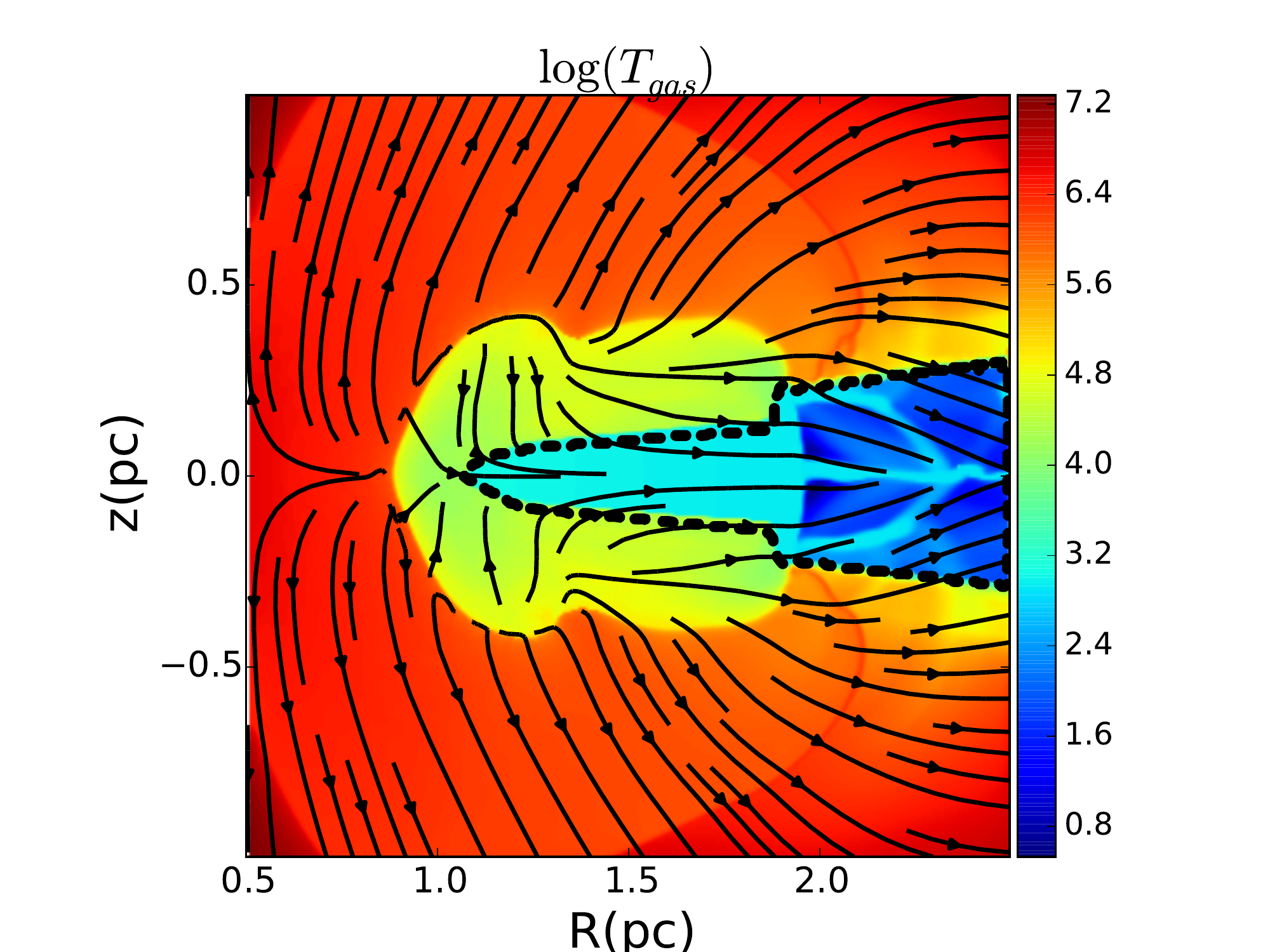}
\caption{
Color plot of the gas temperature, $\log T$ in K for the model with $L= 0.05 L_{\rm edd}$,
at $6.78 \cdot 10^4$ yr.
Axes: horizontal: $z$: distance from equatorial plane in parsecs; $R$: distance from the BH in parsecs;
Inside the dashed contour the dust survives (see text in
Section \ref{ModelL0_05} for discussion).
}\label{figGam005Temp}
\end{figure}

\subsection{Model with $L = 0.05\, L_{\rm edd}$ }\label{ModelL0_05}
 
At larger Eddington ratio radiation pressure plays a stronger dynamical role.
This is illustrated in Figure \ref{figGam005} which shows a sequence of time snapshots
from the evolution of the distribution of the density for $L = 0.05\,L_{\rm edd}$. 

Comparison of Figure \ref{figGam005} with the previous figure shows that the increased illumination
decreases the timescales for all radiation-related processes.  The evaporative flow is established earlier
and is stronger, as is the later evaporation of the disk.  There is no net inflow of the cold thin
disk core; rather there is net outflow of this material starting from the earliest times shown.
This is due in part to the effects of infrared radiation pressure, which disperses the disk and
prevents the establishment of a viscous-dominated inflow, and also to evaporation owing to the
penetration of X-rays into the disk interior.  Figure \ref{figGam005} shows that, at early times,
the dispersal of the disk is strongly affected by radial flow in the disk plane.  This is due to
IR radiation pressure which tends to act in the direction opposite to maximum temperature gradient.
Owing to the presence of warm or hot ($\geq 10^4$K) gas both above and below the disk plane, the
maximum gradient is in the radial direction.  

The maximum velocity is observed in the hot flow where it approaches
$\simeq 700 \, {\rm km\, s^{-1}}$ in the radial, and $\simeq 300 \, {\rm km\, s^{-1}}$ in the vertical
direction (Figure \ref{figGam005}, upper left). 
At later times, in all three other time-steps,  the velocities are in the range $v_r\simeq 90-120\, {\rm km\, s^{-1}}$ and $v_z\simeq 340-400\, {\rm km\, s^{-1}}$.
Following Paper III we calculate the average bulk velocity of the flow $\displaystyle \langle v \rangle = {\int_{V}\,\rho v 
dV} / {\int_{V}\, \rho dV}$, where $V$ is the total or partial volume occupied by the flow. 
The analysis indicates that on average the most massive wind is relatively slow: $\langle v \rangle\simeq 85\, {\rm km\, s^{-1}}$ for the first snapshot shown, $11\, {\rm km\, s^{-1}}$ for the second, and 
$10\, {\rm km\, s^{-1}}$ and $14\, {\rm km\, s^{-1}}$ for the rest.
These velocities are close to quasi-stationary values,
and in accord with the evolution of morphology. This shows similar shapes for the torus at the last three time snapshots. 

The gas temperature in the $\Gamma$=0.05 model is shown in Figure \ref{figGam005Temp}, at the time 6.8 $\times 10^4$ yrs corresponding
to the first panel in Figure \ref{figGam005}.   Also shown are velocity streamlines and the region where dust survives.
One can see that the cold gas is mostly  accumulated within an equatorial disk.
The streamlines illustrate the radial nature of the IR-driven part of the flow, originating in the torus interior.
The effects of wind-compression of the disk at large radii $\geq$2 pc are also apparent.
A shock-like structure at $\simeq$1.8 pc is associated with the radial flow encountering the boundary of the initial torus.
The apparently geometrically thick part of such disk at $R\simeq 2$pc is mostly a relic left from the original torus that 
has been compressed by the hot gas. In its thinnest part, the equatorial disk, shown in Figure \ref{figGam005Temp} and Figure \ref{figGam005} is under-resolved, spanning only $5-10$ grid cells in height.

The dust distribution for the configuration in Figure \ref{figGam005Temp} coincides with the distribution of the 
cold gas shown in blue color.  As we noted before, in our simulations we do not have dust as a separate component, but rather take into account the enhanced opacity due to the presence of dust. That is done whenever
$T_{\rm gas}<T_{\rm sub}$. The corresponding region where dust survives is shown in Figure \ref{figGam005Temp} inside the dashed contour.

At late times the disk at distances greater than 1.2 pc is dispersed into a hot cloud which has a small net 
speed, in either the vertical or the radial direction.  This gas resembles a failed wind, in which
cooling or weak illumination prevent the
gas from attaining a temperature capable of escaping the system.

\subsection{Models with $L= 0.1\, L_{\rm edd}$: evaporative wind}

\begin{figure}[htp]
\includegraphics[width=500pt]{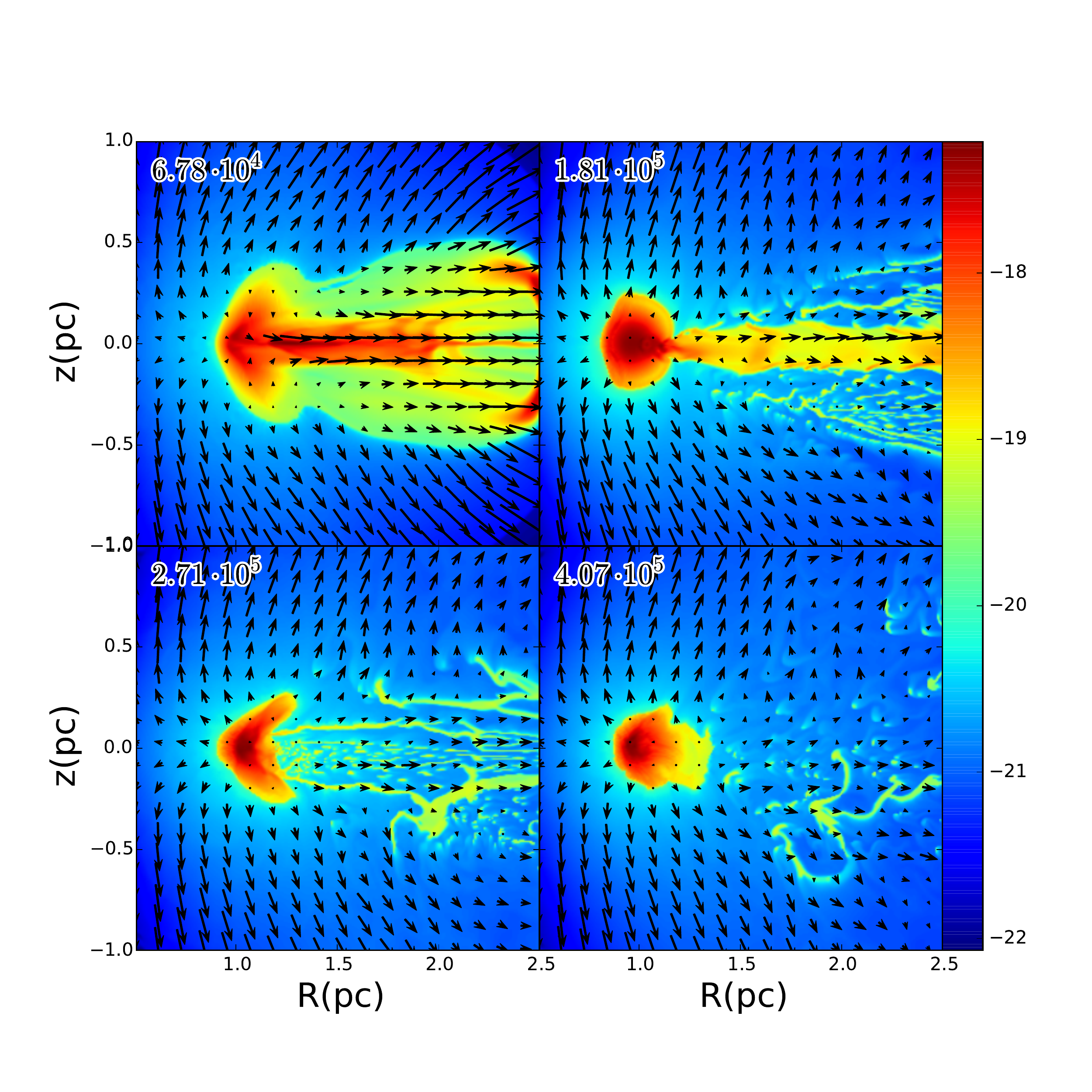}
\caption{
As in Figure 1 but for $L= 0.1 L_{\rm edd}$.
}\label{figGam01}
\end{figure}

The behavior of models with larger luminosity resemble that of the $0.05\, L_{\rm edd}$ model except that step (ii)
of the outline, i.e. the effect of internal IR from absorbed X-rays, is much more rapid and apparent
at short times.
In Papers II and III we found that when luminosity approaches $\simeq 0.1\,L_{\rm edd}$ radiation pressure can disrupt or reverse the accretion flow.  
If so, a key assumption used in those papers, the existence of a thin accretion disk 
situated at the equator which is unaffected by preheating,  may be invalid; here we can test whether similar results are
found without requiring a thin disk boundary condition.
Figure \ref{figGam01} shows the evolution of the torus illuminated by $0.1\, L_{\rm edd}$  from the onset of the simulations (upper left) towards
a complete disruption of the torus into filaments and clouds (lower right). 
The torus is continuously evaporating, replenishing the warm gas.  
The maximum velocity is in the range $\simeq 135-255 \, {\rm km\, s^{-1}}$ for the radial velocity, and in 
$\simeq 350-400 \, {\rm km\, s^{-1}}$ for the vertical one. This approximately equals to
$1.6-2\,V_{\rm esc, 1pc}$, where $V_{\rm esc, 1pc} \simeq 210 \, {\rm km\, s^{-1}}$ is the 
escape velocity at 1pc.

\begin{figure}[htp]
\includegraphics[width= 450pt]{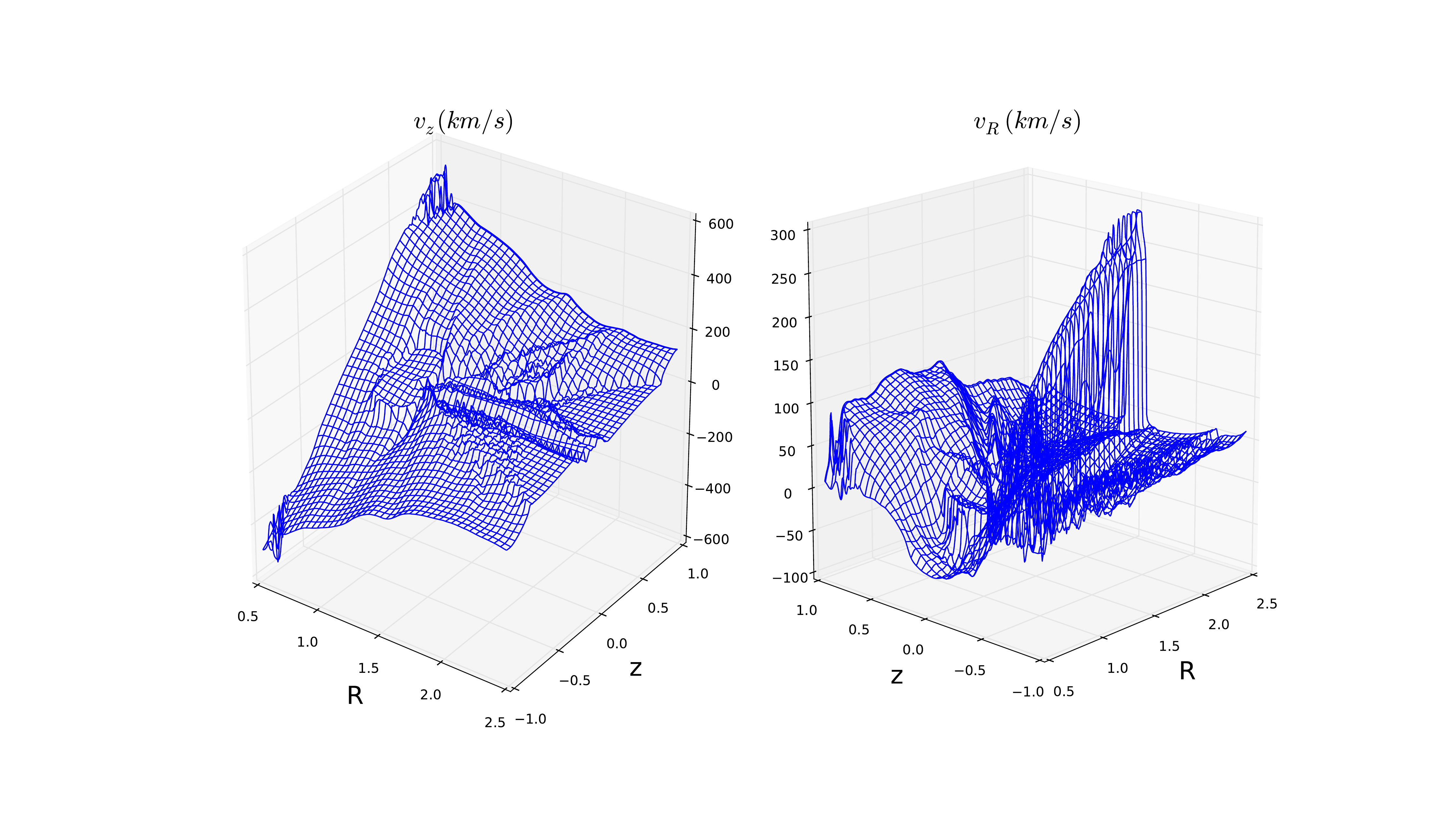}
\caption{ The surface plot of the $v_z$ and $v_R$ velocity components. Model is shown for $L = 0.1L_{\rm edd}$, 
at $t=1.8\times 10^5$yrs. Axes: R: distance from the BH in parsecs; z: distance from the equatorial plane in parsecs.
}\label{wire1}
\end{figure}

Except for the first time snapshot shown, 
$v_z \gtrsim v_r$. The phase of an intense radiation-driven wind lasts for about $10^5$yrs.
Compared with the results for $0.05\, L_{\rm edd}$ , the strong radial flow in the disk plane is
established much earlier and transports material  more rapidly.
At this stage,  $\langle v \rangle\simeq 105\, {\rm km\, s^{-1}}$. As the wind becomes progressively 
more thermally-driven, the mass-loading declines and so does velocity, to $\langle v \rangle \simeq 8-15\, {\rm km\, s^{-1}}$
An equatorial disk-like outflow persists for approximately another $2\times 10^5$yrs. 
This outflow is clearly seen in Figure \ref{wire1}, where polar distributions of $v_z$ and $v_r$ are shown. 
In this figure, an accretion flow can also be recognized as a region of negative radial velocity.
In this disk where $v_r\gg v_z$, both radiation pressure and gas pressure are important, 
and the majority of the radial radiation-driven flow is formed. 
The left panel of Figure \ref{wire1} also shows a fast thermal wind.

At later times ($\geq 2 \times 10^5$ yrs), the part of the wind that has been shielded behind the dense torus evolves
into a disrupted network of over-dense filaments. This  further evolves
into a clumpy structure with total mass $\simeq 3\times10^{3}\,M_\odot$.The results of this simulation are in accord with results
from Paper I: at $L \simeq  0.1-0.3\, L_{\rm  edd}$ the external illumination  significantly affects the flow to the extent that the interior of
the viscous disk is affected.
Radiation pressure on dust can either prevent accretion or it can reduce the accretion rate by squeezing the inflow into a self-shielding, thin equatorial disk. 

\subsection{Models with $L = 0.3\, L_{\rm edd}$ inflow-outflow accretion.}

\begin{figure}[htp]
\includegraphics[width=500pt]{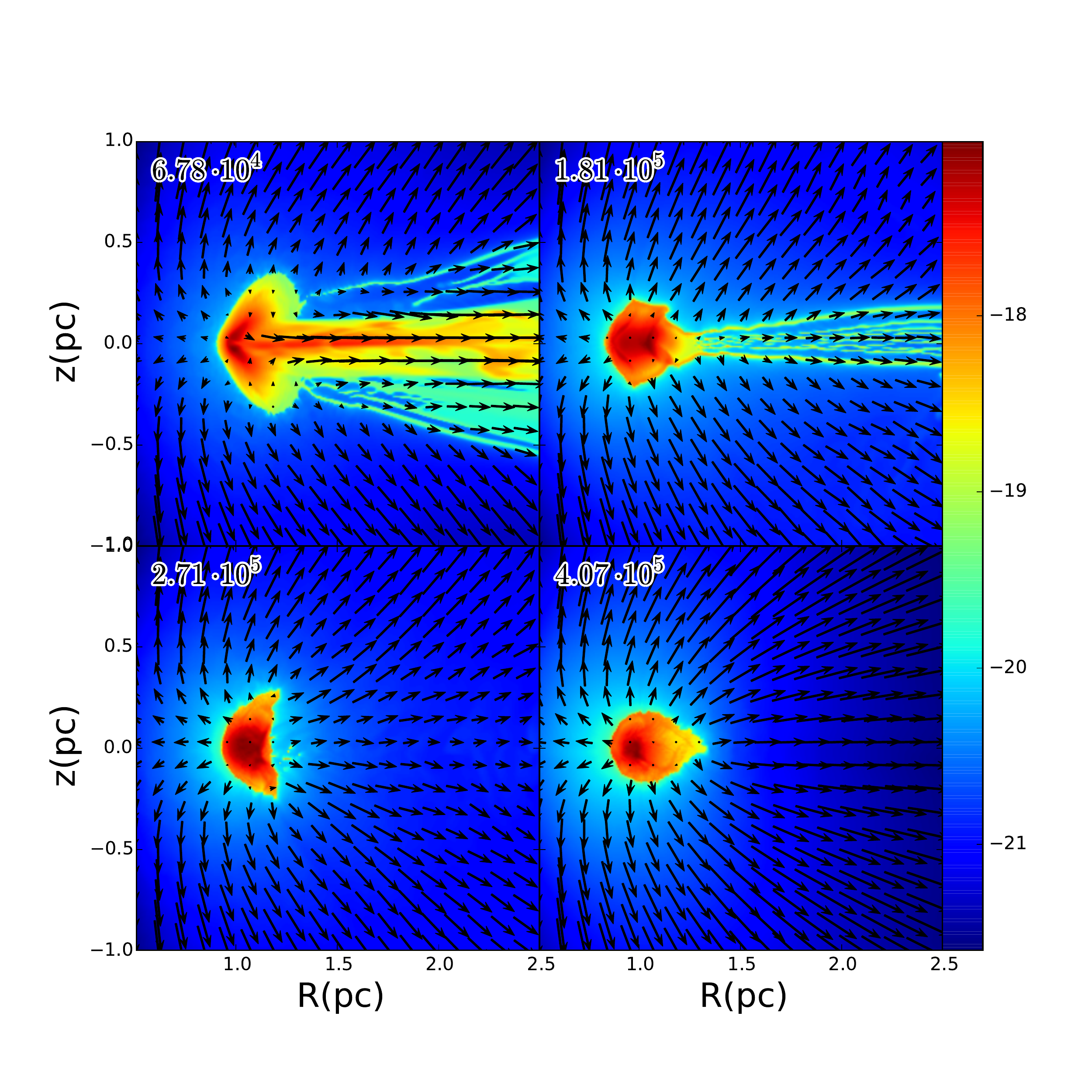}
\caption{
As in Figure 1 but for
$L= 0.3 L_{\rm edd}$.
}\label{figGam03}
\end{figure}

Figure \ref{figGam03} shows a series of time snapshots from a simulation for $L=0.3\, L_{\rm edd}$. Time steps shown are the same as in the three previous models.
The strong illumination in this model removes much of the complexity seen in the previous models.
At $t \lesssim 2 \cdot 10^5$yr a complex tail is seen in the plots, at large radii outside the torus. This is likely a result of thermal instability. The disk behind the denser torus has a layered, sandwich-like structure: cooler and denser layers are intermittent with 
less dense and hotter ones. The dynamical structure is similar to a previous model, depicted in 
Figure \ref{wire1}.
At late times ($\geq 2 \times 10^5$ yrs) there is no evidence for moderate temperature ($\geq 10^4$K) gas
outside the torus; the radiation is strong enough to drive all evaporated material to the Compton temperature and
thus to escape.  At these times the torus possesses a fairly constant evaporative wind. 
As expected, this configuration evaporates faster compared to less luminous cases: within a few dynamical times the initial torus is stripped of most of the gas and the only densest core of the original torus is left evaporating
at the rate $0.1-0.3\, M_{\odot}\, {\rm yr}^{-1}$ (Section \ref{MassLossRateAndWindEnergy}).

\subsection{Mass-Accretion rate}\label{AccretionRate}

\begin{figure}[htp]
\includegraphics[width=300pt]{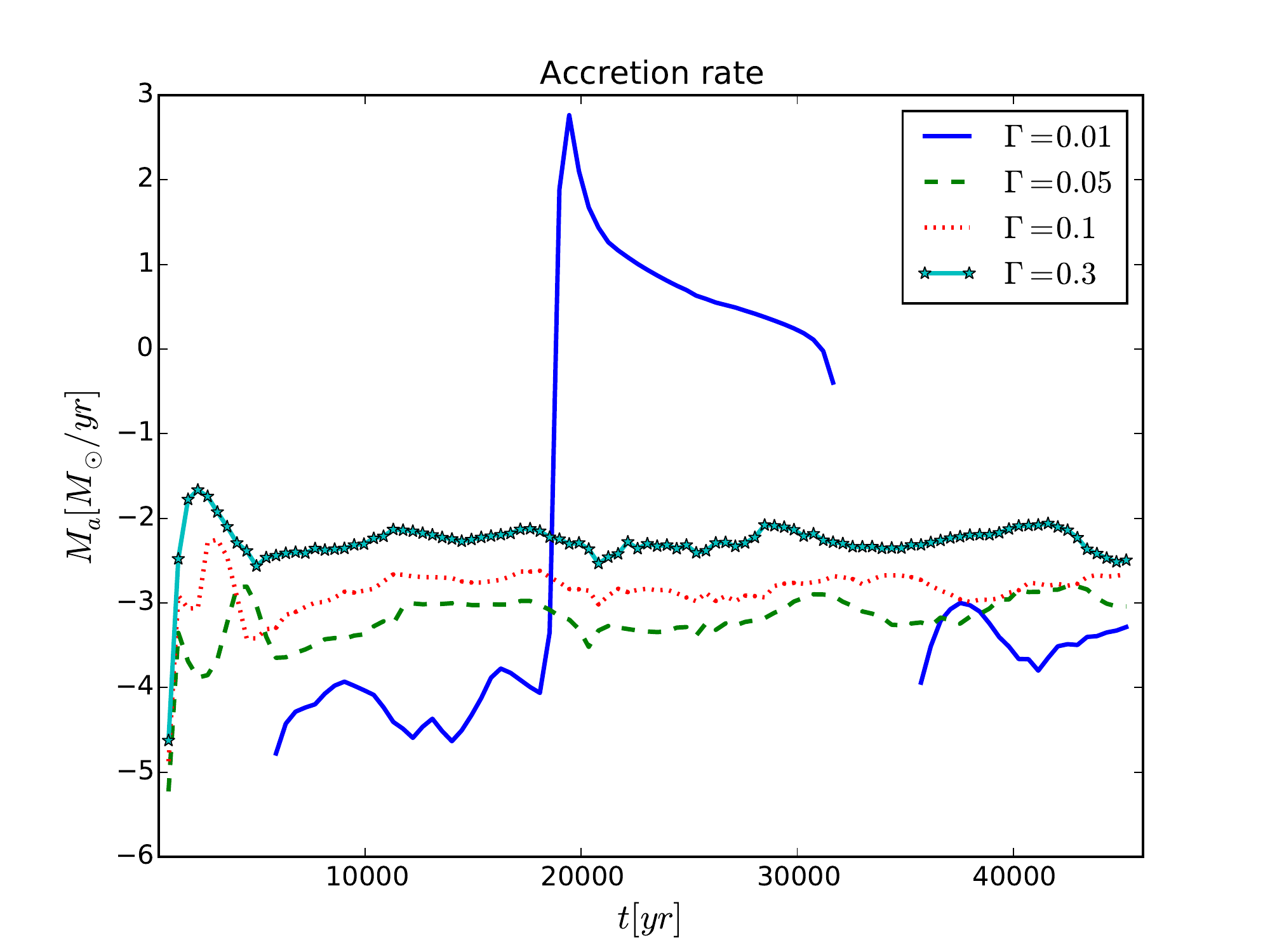}
\caption{
Total accretion rate, ${\dot M}_{\rm a}$ versus time, for 
different values $L/L_{\rm ddd}$. 
Notice, that the model with $\Gamma=0.01$ is the only model that demonstrates a period of disk accretion. After the end of this episode
accretion stops for approximately $3.9 \times 10^3$yrs. 
}\label{figAccretionRate}
\end{figure}

The results in the previous sections show that, when radiation heating is strong (i.e. $L \geq 0.05\, L_{\rm edd}$),
gas that was initially in the disk is dispersed into a hot wind and pushed outward by radiation pressure at
a rate which is faster than the viscosity can diffuse gas inward toward the black hole.  If so, viscous accretion stops. 
Figure \ref{figAccretionRate} shows time-dependent properties of the accretion rate, ${\dot  M}_{\rm a}$,
i.e. the mass flux through the inner boundary of our computational domain.

The model with the lowest radiation input, i.e., $L=0.01 L_{\rm edd}$
demonstrates a period when viscous accretion disk occurs at the equator (c.f. Figure \ref{figGam001}, lower left panel), which corresponds to a rise of the accretion rate by four orders of magnitude within a very short period of time near 2$\times 10^4$ yrs.
The disk thickness can be estimated, assuming radiation pressure support:
$(h/r)\simeq 0.21 \,T_{1500}^2 \sqrt{\frac{ r_{0.5} }{M_7 n_8 } }$, which is in relatively good agreement with Figure \ref{figGam001}.
Other studies of large scale accretion including irradiation by the central AGN found similar behavior:  
despite strong illumination accretion proceeds in the equatorial region because the radiation flux is
significantly reduced  due to the geometrical foreshortening effect and because the equatorial flow is optically thick \citep{Kurosawaproga09}.
Until $t\simeq  1.8  \times 10^4 \,{\rm yr}$ accretion proceeds at the level of $4\times 10^{-4} {\dot M}_{\rm edd}$, where ${\dot M}_{\rm edd}$ is the critical "Eddington" accretion rate, for simplicity taken to be that of a non-rotating BH:
${\dot M}_{\rm edd} = 8.34\times10^{24} M_7 \,{\rm g\,s^{-1} }\simeq 0.13 \,M_7\,M_\odot\,{\rm yr}^{-1}$.
After peaking at
$575 M_{\odot}\, {\rm yr}^{-1} \simeq 4.3\times 10^{3} {\dot M}_{\rm edd}$ at $t\simeq  2  \times 10^4 \,{\rm yr}$  the accretion rate declines and eventually  stops.
This is due to truncation of the accretion disk at $t\simeq  3.2  \times 10^4 \,{\rm yr}$.   That is, due to rapid accretion the disk depletes and the reduced density cannot effectively support screening of its inner parts from overheating and evaporation. 
After some time, hot gas fills the throat of the torus and accretion resumes at much lower level through the capturing of the gas from the wind.

Accretion is interrupted for a period of $3.9  \times 10^3 \,{\rm yrs}$, after which it resumes at the level of 
$10^{-2}-10^{-4}\, M_{\odot}\, {\rm yr}^{-1}$. This second episode of accretion is happening via a
different mechanism of accretion: {\it through capturing of the gas from the wind}.
Note that what we call the 
${\dot M}$ is not the actual rate at which the gas is crossing the innermost stable orbit of the BH.
We only register the gas when it is crossing the boundary of the computational domain. 
The potential energy of the accreted matter will be released in a 
form of radiation - something that we don't address here as it happens much closer to the BH. This will 
correspondingly increase $\Gamma$.

At $L \gtrsim 0.01-0.03 L_{\rm edd}$, the character of the flow changes from net accretion to net outflow.
That is, our models with $L>0.01 L_{\rm edd}$ do not show disk accretion existing for significant amounts of time.
It is possible that more dense models (i.e. higher values of $n_0$) will
allow for the disk to form even at levels of $L$ greater than $0.01 L_{\rm edd}$.
At this level of radiative input, radiation heating and radiation 
pressure hampers the formation of the accretion flow. In a model $L = 0.3 L_{\rm edd}$
it takes $\simeq 2  \times 10^3 \,{\rm yr}$ for the initial configuration to adjust and to achieve quasi-stationary accretion.
All models except the one with $L=0.01 L_{\rm edd}$ accrete at less than the Eddington rate, 
$4.3\times 10^{-3}-10^{-2}{\dot M}_{\rm edd}$.
The model with $L=0.05 L_{\rm edd}$ has the weakest evaporative wind and thus the lowest accretion rate: $10^{-3}-10^{-2}{\dot M}_{\rm edd}$,
while the model with $L=0.3 L_{\rm edd}$ has the highest: $5\times 10^{-2}{\dot M}_{\rm edd}$.

\subsection{Mass-loss rate of the wind}\label{MassLossRateAndWindEnergy}

\begin{figure}[htp]
\includegraphics[width=500pt]{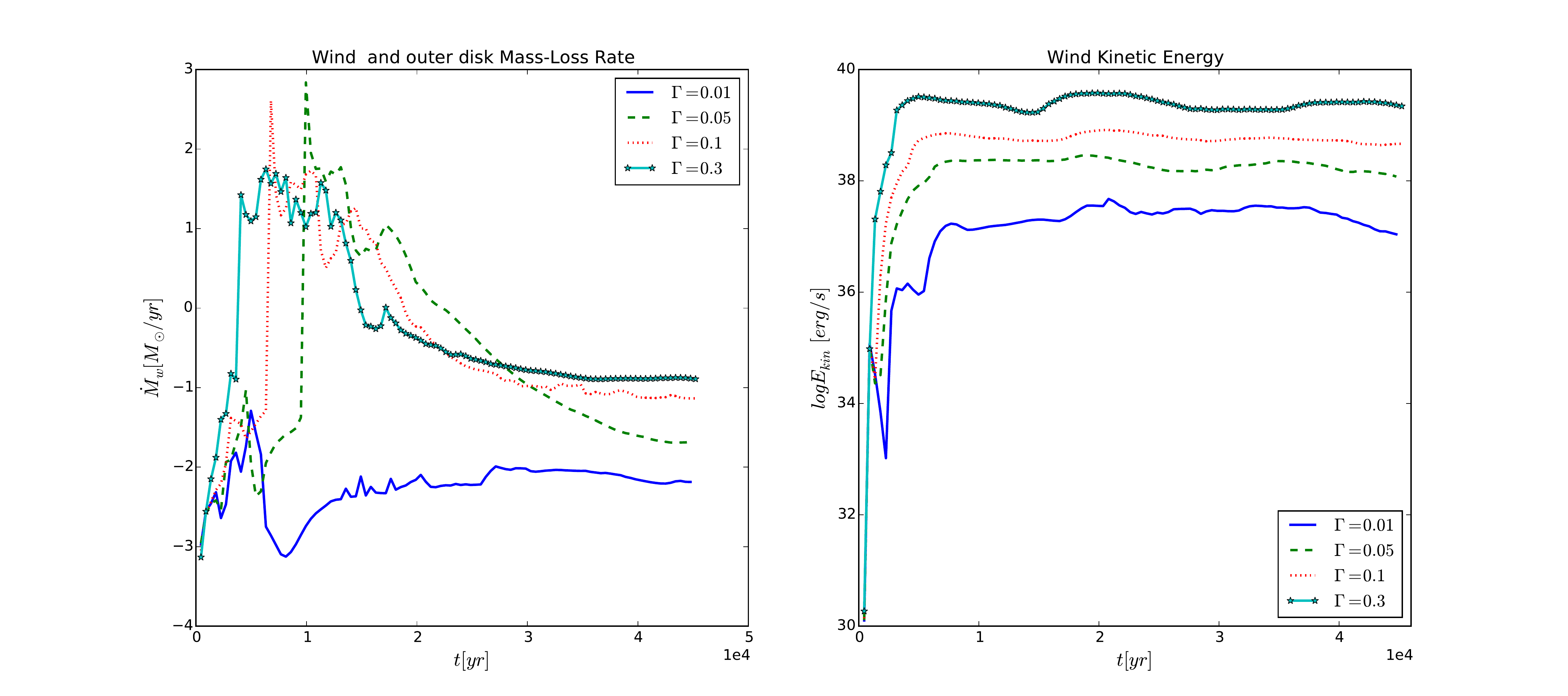}
\caption{
Time dependence of the total wind mass-loss rate , ${\dot M}_{\rm w}$ for fiducial $\Gamma=L/L_{\rm 
edd}$ (left) and the energy of the vertical wind (right).
}\label{figWindRate}
\end{figure}

The wind mass-loss rate, ${\dot M}_{\rm w}$ is calculated at the upper, bottom and at the outer cylindrical 
boundary. Figure \ref{figWindRate} (left) shows the time-dependent properties of  this quantity. 
If the entire torus is expelled within $N_{\rm d}$ dynamical times, with no material accreted, then on average
the wind mass loss rate satisfies
$\langle \dot{M_{\rm w}} \rangle = M_{\rm tor}/ (N_{\rm d} t_0)\simeq 1.04\times 10^2 N_{\rm d}^{-1}\,{\rm M_\odot\,yr^{-1}}$. 
Mass-loss rates observed in the simulations are typically much smaller: In all runs the 
core of the initial torus that contains most of mass survives complete evaporation (c.f. Figure \ref{figGam03}).

From comparing Figure \ref{figAccretionRate} with Figure \ref{figWindRate} (left), the wind mass-loss rate mirrors the accretion rate.
In the model $L=0.01 L_{\rm edd}$ as the inner edge of an accretion disk progresses towards the radiation source, the mass loading from the evaporation increases and produces an increase in ${\dot M}_{\rm w}$. Similar behavior can be noticed for other models.

Models with higher $\Gamma$ have correspondingly higher  ${\dot M}_{\rm w}$: 
For $L=0.01 L_{\rm edd}$ after peaking at ${\dot M}_{\rm w}\simeq   4.7\times 10^{-2}  M_{\odot}\, {\rm yr}^{-1}$,
${\dot M}_{\rm w}$ settles at the level of $10^{-2}  M_{\odot}\, {\rm yr}^{-1}$.
Models with higher values of $\Gamma$
peak at ${\dot M}_{\rm w}\simeq  30 M_{\odot}\, {\rm yr}^{-1}$ gradually reducing to lower values at later times. The peak is associated with the excretion disk and the corresponding strong radial outward motion of the gas
(Figure \ref{wire1}, right panel). Without the outer boundary,  the mass-loss rate would be $4.5\times 10^{-3} M_{\odot}\, {\rm yr}^{-1}$ for $\Gamma = 0.01$, gradually 
increasing to  $0.16 M_{\odot}\, {\rm yr}^{-1}$ for $\Gamma=0.3$.
After an initial period of approximately $(1.5-2) \times 10^4 \,{\rm yrs}$ the mass loss rate settles to a quasi-stationary value.

As mentioned above, at $\Gamma=0.01$ radiation input is not strong enough to prevent the formation of an accretion disk (c.f. Figure \ref{figGam001}).
The accretion flow brings new material in proximity to the
inner boundary of the computational domain where it can be evaporated by radiative heating.
This is a likely explanation of a quicker recovery of  ${\dot M}_{\rm w}$ after $t\simeq 10^{-4}$yr in a 
$\Gamma=0.01$ model.
\subsection{Energy of the wind}\label{WindEnergy}

To assess the dynamical importance of the wind, we calculate the kinetic energy fluxes  at the outer boundaries of the computational domain.
The resultant kinetic luminosity of the wind is $\displaystyle L_{\rm kin}=\int_{\Sigma} \rho v^{3}/2\, d\Sigma$, where the integration is done over the upper and lower boundary that is excluding the excretion disk. It is this 
relatively dilute gas that can be potentially detected as X-ray or UV absorbers.
During initial period of relaxation, thermal output of the wind is $L_{\rm th}\simeq {\rm few}\times L_{\rm kin}$, where 
$L_{\rm th}$ is the thermal output and $L_{\rm kin}$ is the kinetic output. After $\sim 10^3$yrs 
$L_{\rm th}\simeq 0.3-0.6 \times L_{\rm kin}$ for all models.
Within one orbital period the energy output rises to $L_{\rm kin} \simeq 10^{38}{\rm erg\,s^{-1}} $  and stays remarkably constant at this this level until the end of the simulation: $t=10\, t_0 \simeq 4.7\times 10^4\,{\rm yr}$. However, at all times it remains much less than  the total luminosity: $L_{\rm kin}/ L_{\rm accr} \simeq 10^{-5}$. Comparing the two panels of Figure we see that the kinetic output of the $\Gamma=0.01$ model is actually the smallest.  Since the average velocity of the flow is not large, $\sim 10^2$ km/s, 
the dramatic rise of mass loss rate in this model that is happening at 
$t\simeq  2  \times 10^4 \,{\rm yr}$ is not mirrored in the kinetic energy output, c.f. Figure \ref{figWindRate} (right). The energy that the wind carries away is quite low compared to the luminosity of the accretion disk. This is typical for stellar winds, driven by the radiation pressure \citep{LamersCassinelli99}, and was also found in simulations of outflows from accretion flows \citep{KurosawaProgaNagamine2009}.

\subsection{Obscuring properties}\label{ObscuringProperties}

One of key predictions from our models is the dependance the obscuring properties of the flow 
on the inclination angle. 
The apparent torus opening angle is directly relevant to the appearance of an AGN as type I or type II object.
Our previous results suggest that at high radiation input not only radiation pressure but also pressure of the hot wind  are dynamically important. The torus opening angle is determined by  a balance between the gradient of the pressure of gas and radiation from the torus with pressure of the hot gas and radiation pressure from outside the torus.
That is, as a low density, hot gas fills the throat the torus is squeezed towards the equatorial plane.
This, together with the external radiation pressure tends to make the torus aspect ratio, $h/r$  smaller. 

\begin{figure}[htp]
\includegraphics[width=250pt]{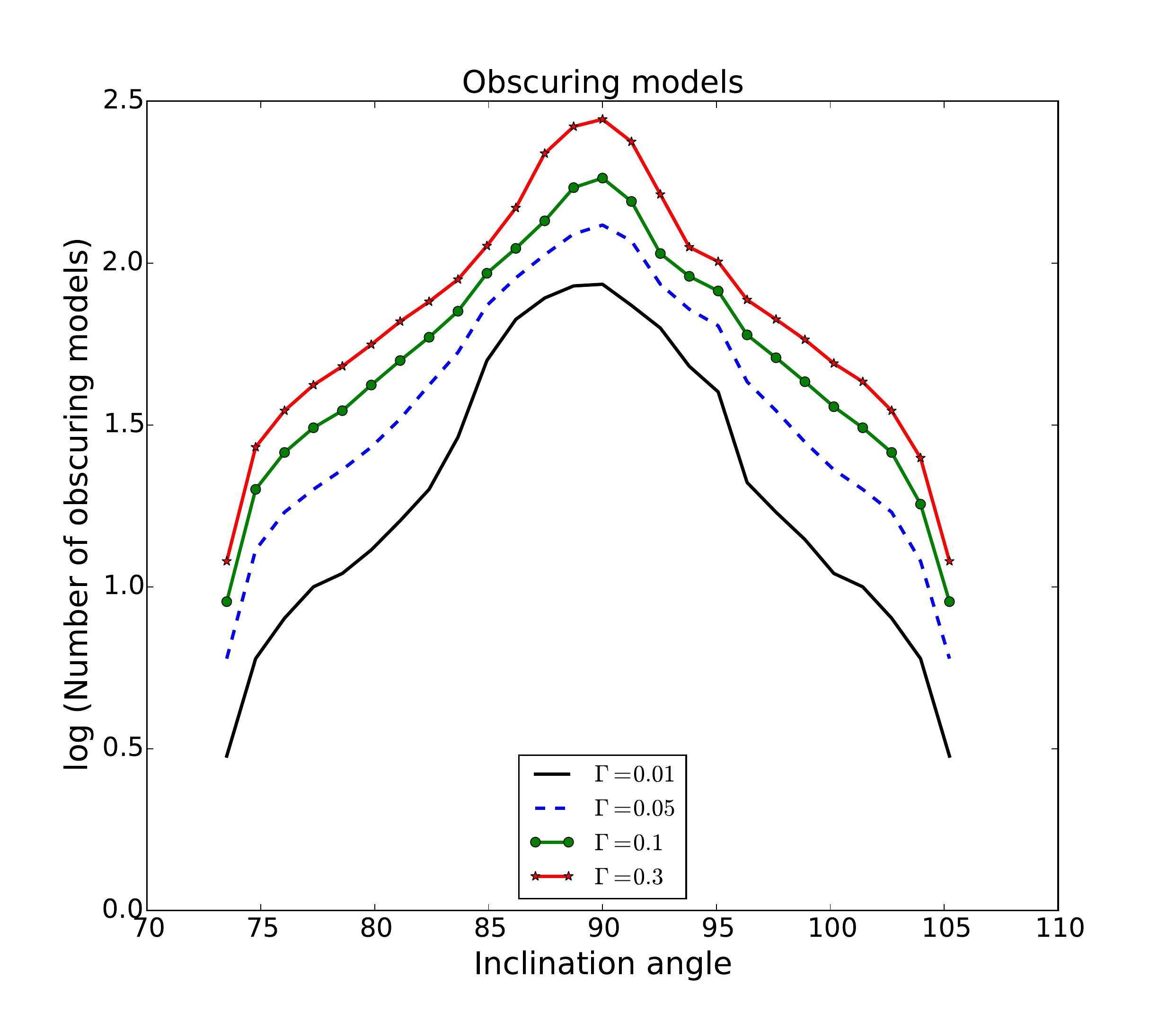}
\caption{
The logarithm of the number of models which at the given inclination have optical depth larger than one. X-axis: the inclination angle in degrees measured from the axis of rotation. Colors indicate different $L/L_{\rm edd}$.
}\label{figObscuration1}
\end{figure}

\begin{figure}[htp]
\includegraphics[width=480pt]{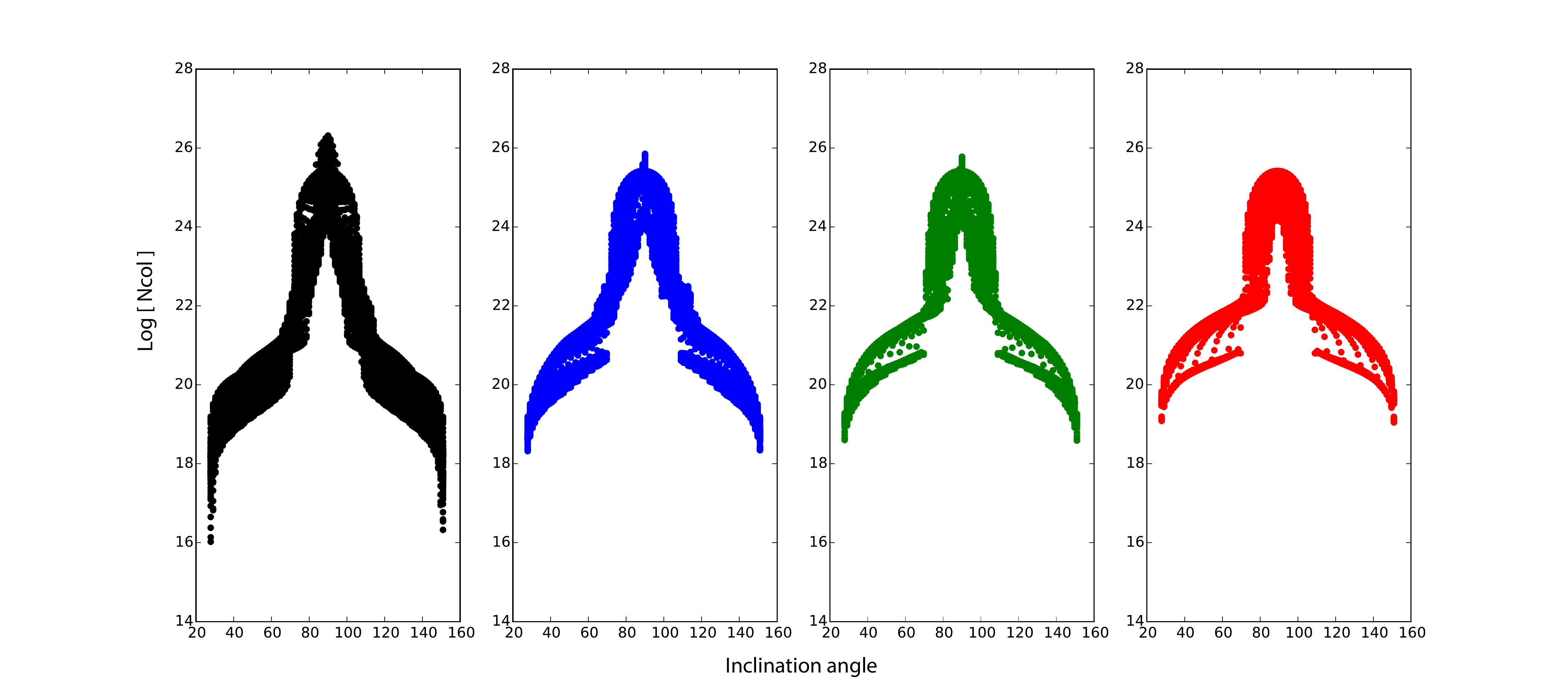}
\caption{
Scatter plot showing the logarithm of the column density, $N_{\rm col}(\text{cm}^{-2}$). Each model is represented by a point.  
Colors indicate different values of $L/L_{\rm edd}$. Models shown from left to right: $\Gamma=0.01$, $0.05$, $0.1$, $0.3$ respectively. 
}\label{figObscuration2}
\end{figure}

As the column density increases towards the equator we are interested in the critical angle, $\theta_{\rm cr}$ at which  the wind becomes opaque. That is when $\tau(\theta_{\rm})\simeq 1$, where  $\tau(\theta)=  \int_{l} \, \kappa_{e}  \rho(l,\theta) \, dl $ is the Thomson optical depth, and  $l$ is measured along a given line of sight, and the viewing angle, $\theta$ is measured from the vertical axis. 
Figure \ref{figObscuration1}, shows the distribution of models with $\tau(\theta)>1$. That is for a given inclination angle, $\theta$ we calculate a total number of model time steps $N_\theta$ which are approximately opaque.
Figure \ref{figObscuration2} show column densities of models at a given inclination. 
Colors correspond to different $\Gamma$.
Each point of a particular color indicates a model at some evolutionary time step. 
Model time span corresponding to the entire span of the time evolution of our simulations, are shown. The total number of models shown, $N_{\rm mod}=100$.

When $\Gamma=0.01$ the episode of disk accretion, leads to a significant depletion of the gas, and thus this set of models has a lower proportion of obscuring models at any inclination. As $\Gamma$ increases, 
no more episodes of disk accretion are observed and larger radiation input leads to larger aspect ratios. 
The distributions of the column densities are broader at higher radiation input, reflecting the importance of 
radiation pressure. The results are that a model with $\Gamma=0.01$ has the smallest and 
a model with $\Gamma=0.3$ has the largest aspect ratio of the torus. 
The region of the torus which is truly Thomson thick subtends a polar angle $\sim 40^o$ at most (for $\Gamma=0.3$);
lower column densities extend to $\sim 80^o$.

\section{Discussion}\label{discussions}


In our previous work we adopted a pre-existing thin accretion disk as a boundary condition. This allowed us to 
study quasi-stationary accretion disk wind solutions but did not allow to address the question of whether accretion can be 
stopped by radiation pressure and heating effects.
Together with these earlier findings, our current results suggest that, qualitatively, obscuration can proceed in accretion 
or outflow mode, depending on the radiative input.  The first scenario comprises a global accretion flow that is slowed, 
puffed up and kept geometrically thick by IR radiation pressure. The second describes a radiation-driven 
obscuring outflow.  
There is also an intermediate mode in which equatorial accretion disk surrounded by slowly in-, or outflowing envelope.
The quantitative boundaries between these scenarios depend on details of heating/cooling, dust formation and 
destruction.  

Some of results presented here may be sensitive to the assumptions regarding microphysics or computational simplifications.
A simple analytical model developed in Paper I of this series predicts that as the temperature of the dusty 
gas reaches
$T_{\rm vir, r}   \simeq 987\,n_7^{1/4} M_7^{1/4} r_{\rm pc}^{-1/4} $K.
the radiation pressure in the optically thick regime becomes larger than gravity and outflow begins.
This is confirmed in the current studies. 
For example, the dynamics of the torus depends primarily on the energy density of infrared radiation and on the ratio of the dust opacity
to the opacity of the fully ionized gas, $\kappa_{\rm d}/\kappa_e$, making the radiation pressure launching of the 
wind depend on how dust is treated. In this paper we adopt relatively simple assumptions about 
the dust formation and destruction.
In later work we plan to examine in more detail the effects of the treatment of dust formation on the hydrodynamics.

It is found that gas accretion is observed throughout the full length of the simulations. However, the details are somewhat surprising: the main path for accretion at $L\gtrsim 0.01\,L_{\rm edd}$ is the accretion from capturing of the thermally driven, evaporative flow. 
Accretion and outflows are both possible at $L\gtrsim 0.01\,L_{\rm edd}$. 
In Paper III we already cast some doubt on the applicability of the accretion disk as a boundary condition due to the finding that the wind mass loss exceed accretion rate needed to produce the required radiative input.
Our current results
suggest that pc-scale disk accretion is likely to be intermittent.

To calculate the X-ray flux we adopt a simple attenuation model neglecting other radiation transfer effects. Effectively, we 
adopt a single scattering approximation when calculating 
local X-ray heating of the gas.
It has been shown \citep{Roth2012} that multiple scattering of X-rays can be important
\citep[e.g.][]{Sim10,Higginbottom14}.  However, no previous work has self-consistently included multiple-scattering
effects with the dynamics.  In this work, we have chosen to focus on the dynamics while neglecting multiple
scattering.

{\mybf 
A lack of the detailed radiation transfer can potentially have an important influence on the results. This is anticipated to improve in the future work.
An additional approximation that we employ is that the frequency integral characterizing the coupling between the
radiation field and the dust opacity is constant.  As a result of reprocessing and expansion 
the IR radiation can be shifted to longer wavelengths in the regions of the flow at the largest radii.
Though it would be interesting to investigate this effect but this is beyond the scope of this paper.}

Other limitations are associated with adopted boundary and initial conditions. One approach is to allow matter to 
constantly enter the domain at the boundary \citep[see e.g.,][]{ProgaBegelman2003,MonikaProga2008,MonikaProga2013}.
However this introduces multiple degrees of freedom such as the 
distribution of matter at the boundary, initial angular momentum of gas, etc.  In this work we  start from a bound torus 
that is close to dynamical equilibrium and study how it reacts to external illumination. 
The limitation of this approach is that we are fundamentally limited by the initial reservoir of gas. We do not allow matter to enter,  and all of the initial torus will be eventually depleted.

The exact characteristics of the accretion flow depend on the assumption of the effective viscosity. 
Limitations of numerical resolution make it unfeasible to capture the 
effects of angular momentum transport in the self-gravitating dusty gas self-consistently. 
We have neglected self-gravity and assumed axisymmetry.
Our results suggest that 
at parsec scales the gas is self-gravitating and can develop strong over-dense filaments and 
clumps on dynamical time-scales. 
Long range gravitational torques from the filaments would lead to angular momentum transport.
Heating of these filaments can provide a negative feedback and make the whole structure hover on the border of 
collapsing into stars, and maintain a Toomre parameter $\Omega_{\rm T} \simeq 1$. 
In the denser part of the accretion disk, such as close to the equator, gravito-thermal instability may generate self-
gravitating turbulence that provide an effective viscosity.  We will explore these processes in future work.

%
%

\section{Conclusions}\label{Conclusions}

We have shown that conversion of external UV and X-ray into IR radiation becomes important at luminosities 
$ \geq 0.01\,L_{\rm edd}$. At characteristic luminosity $L\gtrsim 0.1\,L_{\rm edd}$ 
the effect of radiation heating is profound: both direct X-ray heating and the radiation pressure from IR radiation that is converted from X-rays are altering the dynamics of the torus.

The temperature of the hot gas reaches $10^4-10^6$K: such gas escapes in the form of a fast thermally-driven wind with characteristic velocity of $100-1000\,\kms$; the temperature in the optically thick dusty gas approaches 
$500-1000$K, and the  radiation pressure of infrared photons on dust grains is capable of driving a massive wind with velocities of 
${\rm several}\times 100{\rm \,km\,s^{-1}}$. The typical morphology consists of two kinds of winds: a thermal wind located closer to the BH and the IR-driven wind situated further away. An interface between these two winds typically includes a thin "conversion" layer where X-rays are converted into infrared.  

We find the possibility of three types of gas flows illuminated by intense radiation of AGN:
i) A global accretion flow that is slowed, puffed up and kept geometrically thick by IR radiation pressure; ii) 
An equatorial accretion disk surrounded by slowly in-, or outflowing envelope;
iii) A radiation-driven obscuring outflow. 
At late stages of evolution, the interaction of radiation with the outflowing dense gas creates  a complex structure consisting of compression waves, filaments and clouds. We hypothesize that
strongly compressed gas and the web of shocked filaments and clumps that is found in our $L=0.1\,L_{\rm edd}$ simulations can be possible sites for maser emission. More work is required to study the details of  this regime.

Large amounts of hot, photo-ionized, and dilute gas are routinely observed in our simulations. 
Ram pressure of this component is important in shaping of the cold flow.
Both radiation 
pressure and the dynamical pressure of the hot flow contributes to the "receding torus" effect at high luminosities.
The spectroscopic detection of this gas can be an important task for future missions such as Astro-H and Athena.

\acknowledgements  
This work was supported by NASA under Astrophysics Theory Program grants
10-ATP10-0171  and  NNX11AI96G.

\bibliography{astroph_Dec11}

\end{document}